\documentclass[
  journal=largetwo,
  manuscript=article-type,
  year=2020,
  volume=37,
]{cup-journal}

\usepackage{aas-macros}

\usepackage{amsmath}
\usepackage{amssymb}
\usepackage{upgreek}
\usepackage[nopatch]{microtype}
\usepackage{booktabs}
\usepackage{multirow}
\usepackage{arydshln}
\usepackage{siunitx}
\usepackage{lscape}
\usepackage{subfig}
\usepackage[figuresright]{rotating}
\usepackage{biblatex}
\usepackage[colorlinks=true]{hyperref}
\newcommand*{\fullref}[1]{\hyperref[{#1}]{\nameref*{#1}}}

\newcommand{\g}{G321.3--3.9} 

\newcommand{\fig}{Fig.}
\newcommand{\figs}{Figs.}

\newcommand{\Figs}{Figs.}
\newcommand{\sect}{Section}

\newcommand{\tab}{Table}

\newcommand{\Tab}{Table}
\newcommand{\Tabs}{Tables}

\title{Supernova remnant candidates identified using MWA Galactic Plane Monitoring over $285^{\circ} < l < 70^{\circ}$ and $|b| < 16^{\circ}$}

\author{S. Mantovanini}
\affiliation{International Centre for Radio Astronomy Research, Curtin University, Bentley WA 6102, Australia}
\email[S. Mantovanini]{silvia.mantovanini@postgrad.curtin.edu.au}

\author{N. Hurley-Walker}
\affiliation{International Centre for Radio Astronomy Research, Curtin University, Bentley WA 6102, Australia}

\author{G. E. Anderson}
\affiliation{International Centre for Radio Astronomy Research, Curtin University, Bentley WA 6102, Australia}

\keywords{ISM: individual objects: G16.0+0.7, G39.4-0.0, G39.5+0.4, G299.2-1.5, G306.4+0.1, G309.2-0.1, G310.7-5.4, G317.6+0.9, G321.3-3.9, G322.7+0.1, G324.1+0.0, G324.1-0.2, G328.4-0.1, G332.5-1.2, G332.8-1.5, G333.5+0.0, G335.7+0.9, G336.8-0.6, G341.4-0.2, G352.8-0.3, ISM: supernova remnants, radiation mechanisms: non-thermal}

\bibliography{Biblio}

\begin{document}

\begin{abstract} 
%1. Context.
Observations of Galactic supernova remnants (SNRs) are crucial to understanding supernova explosion mechanisms and their impact on our Galaxy's evolution.  
%2. The problem.
SNRs are usually identified by searching for extended, circular structures in all-sky surveys. 
However, the resolution and sensitivity of any given survey results in selection biases related to the brightness and angular scale of a subset of the total SNR population. As a result, we have only identified 1/3 of the expected number of SNRs in our Galaxy.
%3. How did you solve it?
We used data collected by the Murchison Widefield Array (MWA) to perform a visual search for SNR candidates over $ 285^{\circ} < l < 70^{\circ}$ and $|b| < 16^{\circ}$. We then used the \textit{Widefield Infrared Survey Explorer} to eliminate likely \textsc{Hii} regions from our SNR candidate sample. 
%4. What found.
By exploiting the resolution and sensitivity of MWA data, we have successfully detected 10~new candidates using our proposed method. In addition, our method has also enabled us to detect and verify 10 previously known but unconfirmed candidates.
%5. What that means.
The 20~SNR candidates described in the paper will increase the known SNR population in the Galaxy by 7\%. 
\end{abstract}

\section{Introduction}
A supernova explosion releases $10^{48}$ -- $10^{52} \, \text{ergs}$ of energy and up to tens of hundreds of solar masses into the surrounding environment, contributing to the chemical evolution of the host galaxy. The stellar ejecta left behind results in a supernova remnant \citep[see][for a recent review]{Dubner2015}. Charged particle acceleration, caused by the interactions between the shock wave originated by the intensive explosion and the surrounding medium, induces non-thermal emission across the entire electromagnetic spectrum. 

The duration of the emission changes depending on the frequency; the most energetic particles disperse energy more rapidly during the interactions with the medium, making the lower-frequency radiation emitted by the low-energy particles detectable for longer. This simple mechanism explains why SNRs are usually discovered in the radio band and remain bright for nearly $10^5 \, \text{yrs}$ before becoming indistinguishable from the Interstellar medium (ISM).

Although supernovae deliver large amounts of energy, even becoming brighter than the galaxy that hosts them, throughout human history, we have only seen seven naked-eye supernovae in the Galactic plane: SN~185 \citep[RCW~86; ][]{Clark1977}, SN~386 \citep[$G11.2-0.3$; ][]{Reynolds1994}, SN~1006 \citep{Stephenson1977}, SN~1054 \citep[Crab; ][]{Duyvendak1942}, SN~1181 \citep[3C58; ][]{Stephenson1971}, SN~1572 \citep[Tycho; ][]{Brown1952} and SN~1604 \citep[Kepler; ][]{VandenBergh1977}.
Taking into account the number of stars that result in a supernova in a Milky Way-like galaxy, which is only $1-2$~explosions per century \citep[as discussed in ][]{Tammann1994}, we still expect the presence of nearly $2$~thousand SNRs along the Galactic plane \citep{Frail1994}. However, the number of known remnants is extremely low (a few hundred) compared to this theoretical estimate. 
A few key considerations can explain this:

\begin{itemize}
\item SNRs can be too faint to be detected with the available instruments, particularly for the older population members that have already lost most of their energy.
\item Young SNRs can be very bright but with an angular scale too small to be resolved by the current telescopes.
\item The emission of SNRs can be entangled with the emission of other sources in the background/foreground, making identification difficult.
\end{itemize}

A census of all remnants has been done by \cite{Green2019} and can be found online \footnote{https://www.mrao.cam.ac.uk/surveys/snrs/}. The author updates the catalogue regularly (last in December 2022); to date, it contains 303~confirmed sources and a few hundred candidates, still far from the expected number. Every source is linked with the coordinates, the flux measurement and the spectral index value. As expected, most SNRs have been identified in the radio band, the remaining at X-ray or optical wavelengths. Following the structure of the Green catalogue, \cite{Ferrand2012} created a detailed list of SNRs at high energies, which is regularly kept updated online \footnote{http://snrcat.physics.umanitoba.ca/SNRtable.php}, and summarizes the main properties of each object along with the instruments that were able to detect it. It contains 383~sources, some of which were excluded as confirmed objects from the previously mentioned catalogue due to the absence of a radio counterpart or the lack of strong evidence of non-thermal emission (e.g. spectral index estimates).

New radio surveys have recently provided a promising tool for detecting SNR candidates. A recent example is given by the Evolutionary Map of the Universe \citep[EMU; ][]{Norris2021} and the Polarization Sky Survey of the Universe's Magnetism \citep[POSSUM; ][]{Gaensler2010}, two radio surveys currently observing the Southern sky with the Australian Square Kilometre Array Pathfinder \citep[ASKAP; ][]{Hotan2021} telescope. The EMU/POSSUM Galactic pilot observations, covering an approximate area of $323^{\circ} < l < 330^{\circ}$ and $-4^{\circ} < b < 2^{\circ}$, enabled the identification of 20~SNR candidates, 13 of which were not previously mentioned in other works \citep[see the paper of ][for further details]{Ball2023} due to their faintness and small angular size. 

To fill in the missing population of SNRs, we are using the Murchison Widefield Array \citep[MWA, ][]{Tingay2013,Wayth2018}, a low-frequency radio interferometer located in a remote region of Western Australia. The instrument can cover large areas of the sky ($10^2-10^3$\,sq.deg.) and thanks to this extensive field of view and the wide frequency band $80-300 \, \text{MHz}$, MWA is ideal for searching for SNRs.

The MWA recently performed a revolutionary low-frequency radio survey of the southern skies known as the GaLactic and Extragalactic All-sky MWA \citep[GLEAM; ][]{Wayth2015,Hurley2017} survey, which is very suitable for discovering old SNRs as we expect them to have a low surface brightness and to mainly be detectable at low radio frequencies due to their steep spectra. 
\citet{Hurley2019a} took advantage of GLEAM Galactic Plane images \citep[data release described in ][]{Hurley2019c} and confirmed the SNR nature of two previous candidates and identified 27~new candidates in the sky area between $345^{\circ} < l < 60^{\circ}$, $180^{\circ} < l < 240^{\circ}$. A similar work is being conducted in the area $260^{\circ}<l<340^{\circ}$ by Johnston-Hollitt et al. (in prep.).

This paper is structured as follows. Firstly, \sect~\ref{sec:methodology} introduces the data used and the SNR candidate selection. Then, \sect~\ref{sec:analysis} reports the candidates we have identified with a detailed description of each. Finally, \sect~\ref{sec:discussion} presents a statistical analysis of the sample compared to the SNR population and we detail the advantages of the GPM data in detecting smaller objects and how resolution and sensitivity can be improved in future radio surveys to identify more members of the missing population.

All figures and positions reported in the paper are displayed in J2000 equatorial coordinates unless otherwise specified in the text.

\section{Methodology} \label{sec:methodology}

\subsection{Galactic Plane Monitoring Data}
The primary data used in this paper is the Galactic Plane Monitoring (GPM) campaign (a description of the survey will be presented by Hurley-Walker et al., in prep; and briefly described in the Methods of \citet{Hurley2023}) taken by the MWA across $185 - 215$\,MHz. The campaign was conducted from July to September 2022 with a bi-weekly cadence for the primary objective of identifying transient sources in the Milky Way. This campaign covers $285^{\circ} < l < 70^{\circ}$ and $|b| < 16^{\circ}$, for a total of nearly $4600$\,squared degrees of the southern Galactic plane. Despite being primarily a transient monitoring program, other science goals, such as searching for supernova remnants, can be applied to the same data. 

To generate the image, data were processed as follows: first, the data were flagged for bad antennas, then for RFI using the \textsc{AOFlagger} algorithm \citep{Offringa2012}; next imaging was undertaken with \textsc{WSClean} \citep{Offringa2014} using the GLEAM catalogue as the sky model \citep{Hurley2017}; finally the \textsc{swarp} software \citep{Bertin2002} was used to mosaic the individual snapshots together. This resulted in a final image with a sensitivity of $\simeq 1-2$\;mJy/beam away from bright sources.

Since the GPM campaign searches for transients in Stokes~I images only, here we have generated only total intensity images. Furthermore, as the GPM data was collected using the MWA's longest baseline configuration, the resultant image is sensitive to spatial scales of 45\,arcseconds to 10\,arcminutes, permitting us to estimate the\,200 MHz flux density of extended sources of comparable sizes. As this is only a single frequency, spectral index studies will be the object of future works combining these data with state-of-the-art radio survey data at similar frequencies, such as the upcoming Square Kilometre Array \citep[SKA; ][]{Dewdney2009} map of the plane, which is expected to have a resolution of 2\,arcseconds and sensitivity of $\sim 225 \, \mu \text{Jy/beam}$ along the plane~\footnote{estimated using the SKA sensitivity calculator available at https://www.skao.int/en/science-users/ska-tools/}.

\subsection{Candidate selection}\label{subsec:selection}
To identify candidates in the GPM image of the Galactic plane, we applied the following three steps: 
\begin{itemize}
    \item a visual inspection of the entire longitude range of the GPM image for unidentified ring-shaped structures.
    \item A comparison of these candidates with infrared images to check for thermal counterparts that would rule them out as SNRs.
    \item A comparison of these candidates in other radio surveys to search for similar geometric structures. 
\end{itemize}

First, we overlaid the region files containing all known SNRs classified in \citet{Green2019} on the GPM image and searched for sources outside this sample that show a complete or partial shell-like morphology.

The second step is applied to exclude regions that can be identified as thermal candidates. The morphological structure of SNRs can be confused with \textsc{Hii} regions that abound along the Galactic plane. \textsc{Hii} regions are mostly formed by the ionization of the medium due to the ultraviolet radiation emitted by a newly born star \citep{Condon2016}; the sources are dominated by thermal emission with a power law spectrum (assumed to be $S \propto \nu^{\alpha}$) with a spectral index between $-0.2 < \alpha < +2$. In contrast, we expect a negative spectral index of $-0.7 < \alpha < -0.2$ \citep[for a shell-type remnant as illustrated in ][ nevertheless, exceptions exist]{Dubner2015} for SNRs as their emission is non-thermal synchrotron radiation.
Although we could have the simultaneous presence of thermal and non-thermal emission along the same line of sight due to the richness of sources in the Galactic plane, we excluded all those candidates whose radio shells had a similar structure at infrared wavelengths. To discriminate between regions dominated by thermal or by non-thermal emission, we used three bands of the Widefield Infrared Survey Explorer \citep[AllWISE;][]{Wright2010,Mainzer2011} at $3.4$\;$\mu$m, $12$\;$\mu$m and $22$\;$\mu$m where the morphology of \textsc{Hii} regions is particularly evident. At $\simeq 22$\;$\mu$m the main contribution to the emission of \textsc{Hii} regions is caused by the radiation of hot dust, and it is surrounded by a ring at $\simeq 12$\;$\mu$m caused by Polycyclic Aromatic Hydrocarbon molecules excited by UV radiation from nearby stars \citep[][]{Watson2008}.
 
\begin{table*}[h!]  
\small \centering
\renewcommand{\arraystretch}{1.5}
\begin{tabular}{cccccc}\toprule
& Central frequency & Sensitivity & Angular resolution & Max. angular scale &\\ \multirow{-2}{*}{Survey name} & MHz & mJy/beam & $''$ & $'$ & \multirow{-2}{*}{Reference}\\
\hline \hline   
GPM & 200 & 2 & 45 & 10 & Hurley-Walker et al. in prep.\\
MGPS-2 & 843 & 10 & $45\times45 \csc|\delta|$ & 70 & \cite{Green1999}\\ %shortest baseline of 16m
RACS & 888 & 0.25 & 15 & 15--30 & \cite{Mcconnell2020}\\ %shortest baseline of ASKAP is 22m. For EMU, it can go down to 12m but is not sensitive.
EMU & 944 & 0.025 & 15 & 35 & \cite{Norris2021}\\ 
SMGPS & 1300 & 0.01--0.02 & 8 & 10 &  \cite{Goedhart2024}\\
CHIPASS & 1394.5 & 125 & 864 & N/A & \cite{Calabretta2014}\\ %40\,mK
SPASS & 2303 & 11 & 534 & N/A & \cite{Carretti2019}\\ %9\,mK
Parkes 2.4 & 2400 & 17 & 624 & N/A & \cite{Duncan1997}\\
\bottomrule \end{tabular} 
\caption{Summary of radio surveys in which we searched for our SNR candidates sample. This table includes the acronym name of the survey, the central frequency at which the observations were performed, the sensitivity reached, the angular resolution and the survey description paper.}\label{tab:surveys}
\end{table*}

Finally, surveys at different radio frequencies have been used to compare the emissions of the candidates identified. We report the surveys used throughout the work in \Tab~\ref{tab:surveys}. The variety of sensitivities and angular resolutions that the surveys achieved make it possible to detect fainter objects (as in the case of EMU) or smaller structures, as in the case of the Rapid ASKAP Continuum Survey \citep[RACS; ][]{Mcconnell2020}. Moreover, the new SARAO MeerKAT Galactic Plane Survey \citep[SMGPS; ][]{Goedhart2024} at 1.3\;GHz offers a detailed representation of the fine structure of most of our candidates and a high sensitivity to low surface brightness extended sources. On the other hand, the Molonglo Galactic Plane Survey 2nd Epoch \citep[MGPS-2; ][]{Green2014} at 843\,MHz covers the entire sky south of declination $-30$ degrees, providing a great comparison for all the candidates with latitudes within two degrees of the plane. The Continuum HI Parkes All-Sky Survey \citep[CHIPASS; ][]{Calabretta2014} and the S-band Polarisation All Sky Survey \citep[S-PASS; ][]{Carretti2019} have been used for the analysis of \g. \g\ represents the first important result of this work and has been published in \citet{Mantovanini2024}, in which the authors performed a detailed analysis of the object using data collected by several radio surveys from 200\,MHz to 2300\,MHz and the X-ray survey performed by the eROSITA instrument \citep{Predehl2021}, confirming its nature as an SNR.

The last set of data that we use is the Australia Telescope National Facility pulsar catalogue v2.3.0 \citep[][]{Manchester2005} \footnote{https://www.atnf.csiro.au/research/pulsar/psrcat/} to test whether the remnant candidates that we identified could be associated with a nearby pulsar. A pulsar related to an SNR can yield a better estimate of the shell's distance from Earth and its age. To decide whether a pulsar is associated with our candidates \citep{Kaspi1996}, we take into account various criteria such as the dispersion measure, the pulsar characteristic age (defined as $\dfrac{P}{2\dot{P}}$) and the transverse velocity that must be respectively consistent with a nearby location, the age of an SNR and a common centre of origin. 
The same arguments can be applied to magnetars. We use the McGill Magnetar catalogue \citep[][]{Olausen2014} \footnote{http://www.physics.mcgill.ca/~pulsar/magnetar/main.html} to check whether the candidates are associated with a magnetar located close to the centre of the remnant shell.

\section{Analysis and results}\label{sec:analysis}
Inspecting the image by eye, we found 97~regions with a morphology that resembles an SNR. Comparison to the WISE infrared images eliminated 77~sources as being more luckily \textsc{Hii} regions, reducing the number of SNR candidates to 20. Of the candidates identified, we reviewed the literature and found that ten had been previously listed as SNR candidates, and ten were new candidates. The following sections will provide a detailed description of the sample (see \Tab~\ref{tab:previous} and \Tab~\ref{tab:new}).

We associated each candidate with a category similar to the classification system used in \citet{Brogan2006}: (\textsc{i}) -- the source presents a negative lower limit for the spectral index, which indicates a non-thermal spectrum, it has a complete or partially complete shell morphology, and it does not present an infrared counterpart; (\textsc{ii}) -- we do not have an estimate of the spectral index limit, but the source shows a typical SNR morphology and IR emission does not appear to be in the same sky area; (\textsc{iii}) -- once again, we do not have a limit on the spectral index, the candidate is in a region of great confusion with partial IR contamination. 

As reported in \Tabs~\ref{tab:previous} and \ref{tab:new}, we respectively classified 6, 9 and 5~candidates for each class. 
For each candidate, we provide a morphology description as assessed by visual inspection: ``Shell'' denotes a complete ring displaying limb brightening; ``Partial shell'' denotes a portion $\lesssim 70\%$ of a complete ring is evident in the GPM image; while ``Filled'' denotes an elliptical structure with brightness lacking distinct edges.

Once the list of SNR candidates was refined using the criteria described above, we calculated the flux density of each shell. All the candidates listed have an angular size greater than the maximum scale we could recover; therefore, we treat these measurements as lower limits. 
The flux density has been measured using \textsc{POLYGON\_FLUX}, a software package developed by \citet{Hurley2019a} that permits us to manually draw a polygon around the object to define the region where the flux will be calculated. Since the polygon selection method we employ is subjective, we apply this procedure ten times to each SNR candidate to minimize errors and biases associated with choosing a single polygon. We then applied an interpolated 2D plane to determine the remnant background, which will automatically be subtracted from the selected polygon to get the final flux density measure.
To reduce contamination from other structures just outside the shell, it is possible to draw a second polygon that will exclude all the selected regions from calculating the flux densities and the background. Moreover, it is possible to remove the contribution of point sources within the remnant's shell, which can cause inaccuracies in the flux measurement. The software allows you to click on a source that needs to be removed from the calculation and automatically measures the associated flux to be subtracted from the total flux of the SNR.

Once we have an estimate of the flux density in the GPM data, we refer to the flux density recovery fraction plot reported in \fig~2 of \citet{Hurley2022} to predict how underestimated is the flux density measure. The plot shows the fraction of the measured and model flux densities as a function of the angular scale for sources of different brightness and various cleaning models with a weighting parameter varying from uniform to natural. This provides us with a lower limit at the GPM frequency.
When SMGPS data is available, and the source size falls within the sensitivity of the survey, we estimated the flux density applying the \textsc{POLYGON\_FLUX} software. As \citet{Goedhart2024} detailed, spectral index values tend to be more precise within the middle of the band (approx. 1300\,MHz), where flux density measurements are in good agreement with values found in the literature. While we may miss some flux, we anticipate it to scale with the maximum spatial scale the survey could recover.

Finally, we determine the spectral index of the source by combining the two estimates mentioned above. This spectral index measurement, therefore, provides insight into whether the source could be synchrotron-dominated. If our estimate of the spectral index lower limit falls within the range expected from SNRs, we can confidently conclude that we are dealing with a non-thermal object, validating our detection.

The following two sections will highlight the main characteristics of the candidates that were previously identified in the literature and, in particular, the new candidates we have identified with these data.

\subsection{Previous SNR candidates} \label{sub:previous}
In this section, we illustrate the 10~SNR candidates we have detected with the GPM data already cited by previous works in the literature. One of these candidates corresponds to \g\, which has been confirmed as an SNR thanks to a combined analysis at radio frequencies and X-rays by \citet{Mantovanini2024}. The object has been included in the table and the statistics analysis in \sect~\ref{sec:discussion} for completeness, but further details can be found in the recently published paper. In \Tab~\ref{tab:previous}, we summarize these objects' main properties and reference the first paper that reported them. The region where the candidates reside can be seen in \Figs~\ref{fig:G39.5+0.4}-\ref{fig:G336.8-0.6}; in each figure, the first panel corresponds to the GPM data at 200\,MHz used in this work, the middle panel shows either RACS, EMU or SMGPS data as specified in the caption, and the third panel represents a RGB image constructed using the higher three bands of the WISE survey as described in \sect~\ref{subsec:selection}. 

According to the classification system defined in \sect~\ref{sec:analysis}, six~candidates fall in the first category, three in the second one and only one in the third, meaning we are fairly confident that the sources are remnants, and they do not show a typical \textsc{hii} region morphology in infrared coincident to the radio shell. We cannot accurately estimate the flux densities of these sources due to their faintness and the survey's poor resolution to their angular scales, which is inadequate to resolve the entire structure of the candidates. However, a lower limit is reported in \tab~\ref{tab:previous}.

Most candidates are fairly close to the Galactic plane with $|b| < 1^{\circ}$ except for G299.2$-1.5$, G310.7$-5.4$ and G321.3$-3.9$. Those objects correspond to three of the four largest candidates we have identified in our sample. This fact can be explained by considering that at high Galactic latitudes, the density of the ISM decreases, reducing the rate of energy dispersion; consequently, these sources can expand for longer times and at a higher rate, reaching larger radii.\newline

\begin{centering}
\begin{sidewaystable*}[htbp]
\small \centering
\renewcommand{\arraystretch}{2}
\begin{tabular}{ccccccccccccccc} \toprule
& RA & Dec & $l$ & $b$ & MAJ & MIN & PA & $S_{200\,MHz}$ & & & & S$_{survey}$ & &\\ \multirow{-2}{*}{Name} & J2000 & J2000 & $^{\circ}$ & $^{\circ}$ & $^{\circ}$ & $^{\circ}$ & $^{\circ}$ & Jy & \multirow{-2}{*}{Morphology} & \multirow{-2}{*}{Class} & \multirow{-2}{*}{Survey} & Jy & \multirow{-2}{*}{Ref. $^a$} & \multirow{-2}{*}{Spectral index}\\
\hline \hline 
\hyperref[G39.5+0.4]{G39.5+0.4} & 19:01:56 & +6:04:49 & 39.54 & 0.45 & 0.30 & 0.28 & 302 & > 1.3 & Shell? & \textsc{iii} & GLOSTAR & > 0.024 $\pm$ 0.003 & 1 &  \\ 
\hyperref[G299.2-1.5]{G299.2$-1.5$} & 12:17:29 & $-64$:09:15 & 299.24 & $-1.53$ & 0.50 & 0.50 & 0 & > 4.5 & Filled & \textsc{ii} & MGPS-2 & > 0.31 & 2 & \\
\hyperref[G310.7-5.4]{G310.7$-5.4$} & 14:12:15 & $-67$:03:37 & 310.70 & $-5.42$ & 0.45 & 0.45 & 0 & > 1.1 & Shell & \textsc{ii} & MGPS-2 & > 0.83 & 2 & \\ 
\hyperref[G317.6+0.9]{G317.6+0.9} & 14:47:29 & $-58$:38:49 & 317.55 & 0.89 & 0.58 & 0.44 & 80 & > 1.9 & Partial shell & \textsc{ii} & MGPS-1 & & 3 &\\  
\hyperref[G321.3-3.9]{G321.3$-3.9$} & 15:32:42 & $-60$:51:49 & 321.34 & $-3.89$ & 1.7 & 1.1 & 270 & > 6.1 & Shell & \textsc{i} & Parkes 2.4-GHz & & 4 & -0.8 $\pm$ 0.2\\ 
\hyperref[G322.7+0.1]{G322.7+0.1} & 15:23:54 & $-56$:49:07 & 322.68 & 0.09 & 0.21 & 0.21 & 0 & > 0.7 & Filled & \textsc{i} & MGPS-2 & > 0.1 & 5 & $-1^{+0.1}_{-0.3}$\\ 
\hyperref[G324.1-0.2]{G324.1$-0.2$} & 15:33:33 & $-56$:13:43 & 324.11 & $-0.17$ & 0.14 & 0.14 & 0 & > 0.24 & Filled & \textsc{i} & EMU & > 0.2 & 5 & $< -0.1$\\
\hyperref[G324.1+0.0]{G324.1+0.0} & 15:32:37 & $-56$:02:59 & 324.11 & 0.05 & 0.17 & 0.12 & 302 & > 1.7 & Shell & \textsc{i} & MGPS-2 & 0.6 & 5 & $< -0.6$\\  
\hyperref[G328.4-0.1]{G328.6+0.0} & 15:57:31 & $-53$:21:53 & 328.59 & $-0.03$ & 0.15 & 0.27 & 0 & > 0.9 & Filled \& partial shell & \textsc{i} & SGPS & 3.5 $\pm$ 0.9 & 6 & $-0.75 \pm 0.06$\\
\hyperref[G336.8-0.6]{G336.8$-0.6$} & 16:37:14 & $-47$:58:22 & 336.85 & $-0.54$ & 0.18 & 0.28 & 320 & > 1.6 & Shell & \textsc{i} & MGPS-2 & 0.66 $\pm$ 0.6 & 2 & < $-0.6$\\ 
\bottomrule 
\end{tabular} 
\caption{Properties of previous SNR candidates detected in this work as detailed in subsection \ref{sub:previous}. The table is structured as follows: \textbf{Name} derived from Galactic coordinates via \textit{lll.l} $\pm$ \textit{b.b}; \textbf{Ra, Dec} right ascension and declination in J2000 coordinates; \textbf{l, b} longitude and latitude in Galactic coordinates; \textbf{MAJ, MIN} major and minor axes of the candidate's elliptical shell in degrees; \textbf{PA} is the position angle in degree; \textbf{S$_{200}$} flux density measured at 200 MHz; \textbf{Morphology} of the shell as can be determined by visual inspection; assigned \textbf{Class} (as defined in \sect~\ref{sec:analysis}); \textbf{Survey} name of the survey where the candidate was previously detected; \textbf{S$_{survey}$} is the flux density measure as provided by an additional survey and referenced in the next column; \textbf{Ref} reports the reference number of the work that estimated the flux density in the previous column; \textbf{Spectral index} limit as derived with flux densities from columns 9 and 13 unless otherwise specified.}
\footnote{Publications listing the flux density measure reported in the previous column and used to determine the spectral index: (1) \citet{Dokara2021}, (2) \citet{Green2014}, (3) \citet{Green1999}, (4) \citet{Mantovanini2024}, (5) This work, (6) \citet{McClureGriffiths2001}.}\label{tab:previous}
%(1)\citet{Dokara2021}, (2) \citet{Duncan1997}, (3) \citet{Green2014}, (4) \citet{Green1999}, (5) \citet{Ball2023} and (6)~\citet{McClureGriffiths2001}.}
\end{sidewaystable*}
\end{centering}

In the following, a brief description of each candidate is provided. 

\subsubsection{G39.5+0.4}\label{G39.5+0.4}
\begin{figure*}
\centering
\includegraphics[width=1.0\linewidth]{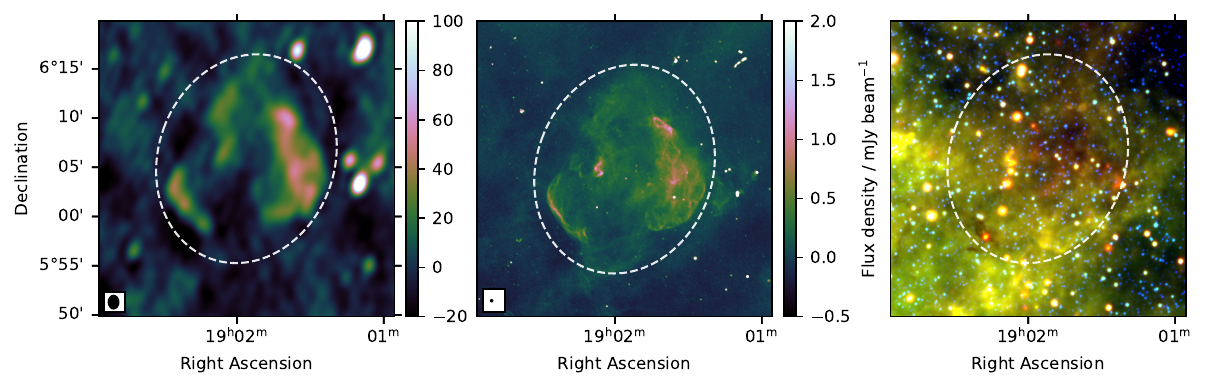}
\caption{Region surrounding \hyperref[G39.5+0.4]{G39.5+0.4} (class \textsc{iii}) as observed by GPM at 200 MHz (left), by SMGPS at 1300 MHz (middle) and by WISE (right) at 22 $\mu$m (R), 12 $\mu$m (G), and 3.4 $\mu$m (B).}
\label{fig:G39.5+0.4}
\end{figure*}

This candidate has been identified by \citet{Dokara2021} using the Karl G. Jansky Very Large Array GLObal view of the STAR formation in the Milky Way (GLOSTAR) survey at 5.4\,GHz as a single arc morphology with a linearly polarised emission with a degree of polarization $0.06 \pm 0.02$. Our detection can better resolve the object's structure (refer to the first panel of \fig~\ref{fig:G39.5+0.4}), making almost the entire shell visible with a $0.30 \times 0.28$ degrees dimension. The authors estimated a lower limit for the flux density at 5.4\,GHz obtaining a value of $0.024 \pm 0.003$\,Jy as the instrument is only sensitive to structures smaller than approximately 1.5\,arcminutes. The object is visible in other radio surveys (GLEAM, RACS, and SMGPS are the most relevant examples), which assess our confidence level in the object. The candidate, as observed by SMGPS at 1300\,MHz is reported in the middle panel of  \fig~\ref{fig:G39.5+0.4}, while the right panel shows the presence of some IR emission in the same region of the shell's candidate, which does not share the same morphology but it is hard to clarify whether the two emission contribution are related to each other with the current data available. Further observations are necessary to disentangle the synchrotron emission from a possible \textsc{hii} region. The absence of a spectral index limit and the contamination from thermal emission made us classify this candidate as a class~\textsc{iii}.  

\subsubsection{G299.2$-1.5$}\label{G299.2-1.5}
\begin{figure*}
\centering
\includegraphics[width=1.0\linewidth]{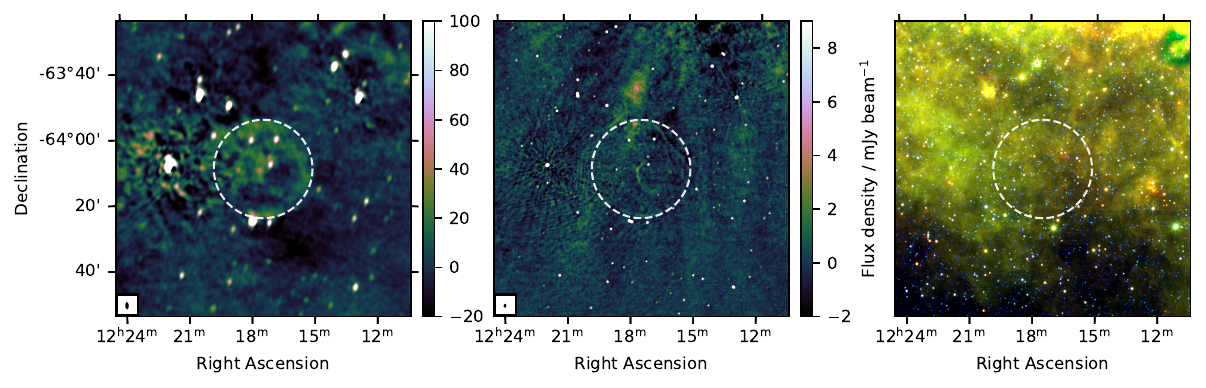}
\caption{Region surrounding \hyperref[G299.2-1.5]{G299.2$-1.5$} (class \textsc{ii}) as observed by GPM at 200 MHz (left), by RACS at 888 MHz (middle) and by WISE (right) at 22 $\mu$m (R), 12 $\mu$m (G), and 3.4 $\mu$m (B).}
\label{fig:G299.2-1.5}
\end{figure*}

This SNR candidate represents one of the largest objects in our sample; it has been first identified by \citet{Duncan1997} as a circular object with a 35\,arcminutes diameter, and later on by \citet{Green2014} as a faint but complete circular shell at 843\,MHz. Due to the instrument limitation, the authors could only be able to provide a lower limit of the flux density and estimated it to be greater than 0.31\,Jy. We determined that the integrated source flux should be greater than 4.5\,Jy in agreement with a falling spectrum of an SNR object. Albeit the WISE RGB image (see panel three of \fig~\ref{fig:G299.2-1.5}) does not show the presence of a \textsc{hii} region structure in the same sky region, we have only classified the source as class \textsc{ii} due to its large angular size which made us difficult to set some constraints on the spectral index. The structure of the object as seen by the GPM at 200\,MHz and RACS at 888\,MHz is shown in the left and middle panels of \fig~\ref{fig:G299.2-1.5}. The candidate has also been detected in Johnston-Hollitt et al. (in prep.).

\subsubsection{G310.7$-5.4$}\label{G310.7-5.4}
\begin{figure*}
\centering
\includegraphics[width=1.0\linewidth]{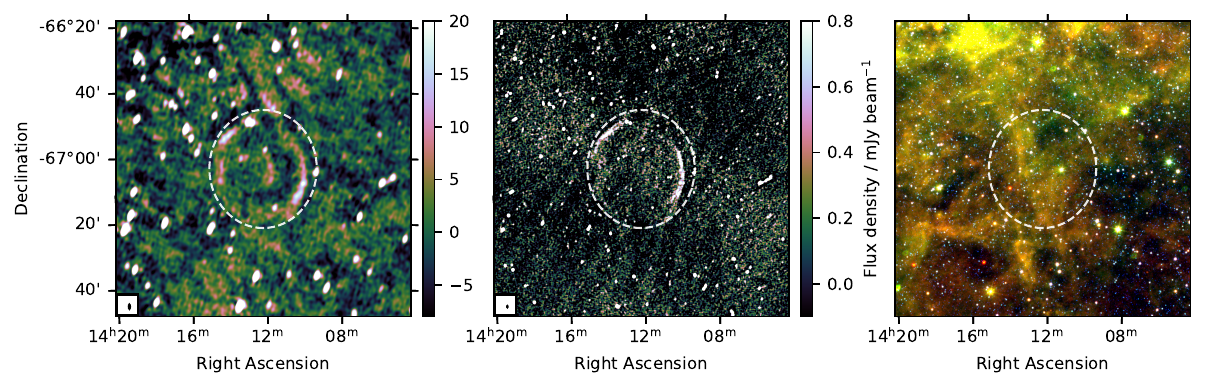}
\caption{Region surrounding \hyperref[G310.7-5.4]{G310.7$-5.4$} (class \textsc{ii}) as observed by GPM at 200 MHz (left), by RACS at 888 MHz (middle) and by WISE (right) at 22 $\mu$m (R), 12 $\mu$m (G), and 3.4 $\mu$m (B).}
\label{fig:G310.7-5.4}
\end{figure*}

The SNR candidate G310.7$-5.4$ was discovered by \citet{Green2014} and described as a complete circular shell with a diameter of $31 \times 29$\,arcminutes$^2$. Similarly, we observe two prominent arcs composing a round shell of 27\,arcminutes in diameter, making it one of the more extensive objects in the sample (see left and middle panel of \ref{fig:G310.7-5.4}); inspecting the GPM image, we noticed the presence of a fainter arc within the primary shell, suggesting the possible presence of a reverse shock. The WISE RGB image does not show any IR emission associated with it, as illustrated in the right panel of \fig~\ref{fig:G310.7-5.4}. The first detection could not fully resolve the object's structure; once again, we can only provide a lower limit on the flux density, corresponding to 1.1\,Jy. We denoted the object as class \textsc{ii} according to our classification scheme. 

\subsubsection{G317.6+0.9}\label{G317.6+0.9} 
\begin{figure*}
\centering
\includegraphics[width=1.0\linewidth]{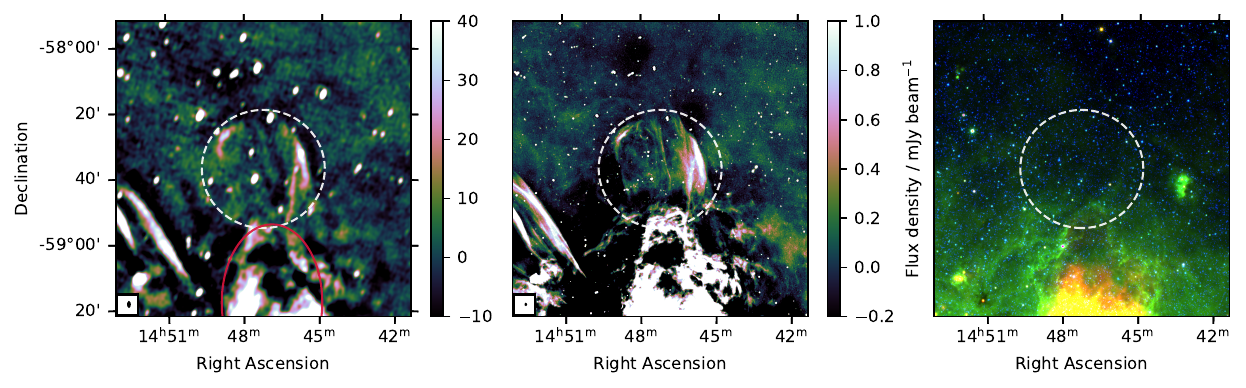}
\caption{Region surrounding \hyperref[G317.6+0.9]{G317.6+0.9} (class \textsc{ii}) as observed by GPM at 200 MHz (left), by EMU at 944 MHz (middle) and by WISE (right) at 22 $\mu$m (R), 12 $\mu$m (G), and 3.4 $\mu$m (B). The red ellipse surrounds thermal contributions in the candidate sky area.} 
\label{fig:G317.6+0.9}
\end{figure*}

This source is listed as a potential SNR candidate by \citet{Green1999} but has not been confirmed. We observe a partial shell structure in the GPM image (left panel of  \fig~\ref{fig:G317.6+0.9}) with a diameter of $0.58 \times 0.44$\,degrees$^2$. The EMU survey better resolves the morphology as illustrated in the middle panel of \fig~\ref{fig:G317.6+0.9}, where it is possible to distinguish a filament structure. The candidate's dimensions are too big and can not be resolved by the GPM campaign. Therefore, we could only estimate a flux density lower limit, which corresponds to 1.9\,Jy. The third panel of the figure instead shows the absence of an infrared structure within the radio shell, which enables the classification of the object as part of class \textsc{ii}. The candidate has also been listed in Johnston-Hollitt et al. (in prep.).

%\subsubsection{G321.3-3.9}\label{G321.3-3.9} %M
%G321.3-3.9 represents the biggest candidate in this work, with minor and major diameters of 1.1 and 1.7\,degrees, respectively. A detailed analysis of the object \citep[see ][]{Mantovanini2024} using data collected by several radio surveys from 200\,MHz to 2300\,MHz has been performed. Using data collected by the single-dish surveys CHIPASS and S-PASS, the authors calculated a spectral index of $\alpha \simeq -0.8 \pm 0.2$ consistent with the expected non-thermal emission from an SNR. The extended structure of this source also appeared in the all-sky survey performed by the X-ray telescope eROSITA \citep{Predehl2021}, which enabled the authors to estimate a column absorption of the order of $N_{\text{H}} \sim 10^{21}$ cm$^{-2}$ thanks to the application of spectral models to the photon statistics. These measurements enabled the authors to confirm the classification of the object as an SNR. The candidate has also been detected in Johnston-Hollitt et al. (in prep.).

%\begin{figure*}
%\centering
%\includegraphics[width=1.0\linewidth]{SNR_G321.3-3.9.pdf}
%\caption{Region surrounding \hyperref[G321.3-3.9]{G321.3-3.9} (class \textsc{i}) as observed by GPM at 200 MHz (left), by EMU at 944 MHz (middle) and by WISE (right) at 22 $\mu$m (R), 12 $\mu$m (G), and 3.4 $\mu$m (B).}
%\label{fig:G321.3-3.9}
%\end{figure*}

\subsubsection{G322.7+0.1}\label{G322.7+0.1} %M
\begin{figure*}
\centering
\includegraphics[width=1.0\linewidth]{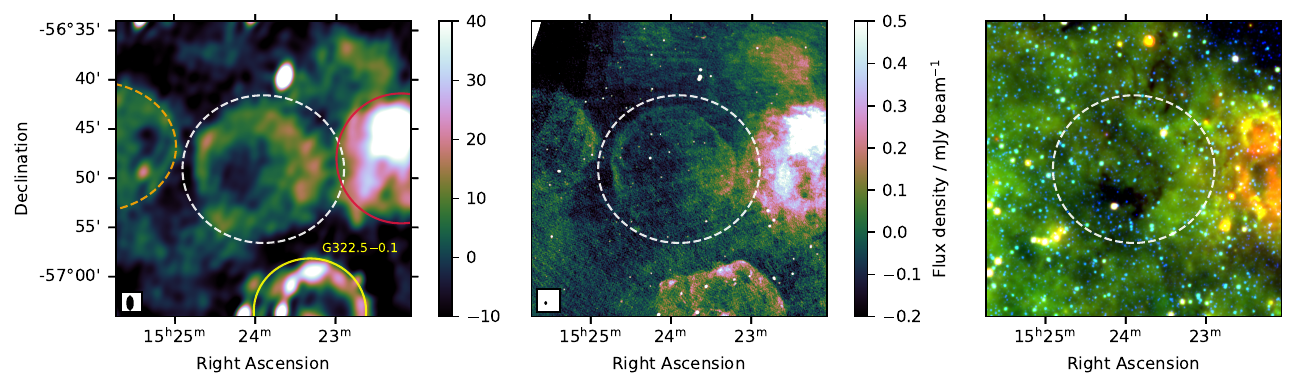}
\caption{Region surrounding \hyperref[G322.7+0.1]{G322.7+0.1} (class \textsc{i}) as observed by GPM at 200 MHz (left), by SMGPS at 1300 MHz (middle) and by WISE (right) at 22 $\mu$m (R), 12 $\mu$m (G), and 3.4 $\mu$m (B). The known remnant G322.5-0.1 is highlighted in yellow; the orange dashed line goes around an SNR candidate, while the red ellipse surrounds thermal contributions in the candidate sky area.}
\label{fig:G322.7+0.1}
\end{figure*}

The object was first proposed as a potential remnant by \citet{Whiteoak1996}, but an image of the object was only provided two decades later by \citet{Green2014}, where it appeared as an extremely faint and circular shell. A similar structure is observed in the GPM image (see left panel of \fig~\ref{fig:G322.7+0.1}) with a radius of 0.1\,degrees. \citet{Green2014} estimated a flux density of 0.29\,Jy at 843\,MHz with an error between 5 and 10\%; considering a mean Galactic SNR spectral index of -0.51 \citep{Ranasinghe2023}, the source's flux at the GPM frequency should vary between 0.5 and 0.7\,Jy, in accordance with the lower limit provided by us (\Tab~\ref{tab:previous}). A limit on the spectral index can be obtained using GPM and MGPS-2 results, which converge in $\alpha < -0.5$. A better estimate is provided by combining SMGPS (middle panel of \fig~\ref{fig:G322.7+0.1}) and GPM data; using \textsc{POLYGON\_FLUX}, we estimated a flux density of 0.1\,Jy at 1300\,MHz which lead to a spectral index of $-1^{+0.1}_{-0.3}$, where the uncertainties take into consideration the fraction of the flux densities that we might be missing from the surveys involved in the calculation. The infrared image in the same sky area does not show any \textsc{hii} region-like structure (right panel of \fig~\ref{fig:G322.7+0.1}), increasing our confidence level in this candidate, which has been classified as class~\textsc{i}. The candidate has also been detected in Johnston-Hollitt et al. (in prep.).

\subsubsection{G324.1$-0.2$}\label{G324.1-0.2} 
\begin{figure*}
\centering
\includegraphics[width=1.0\linewidth]{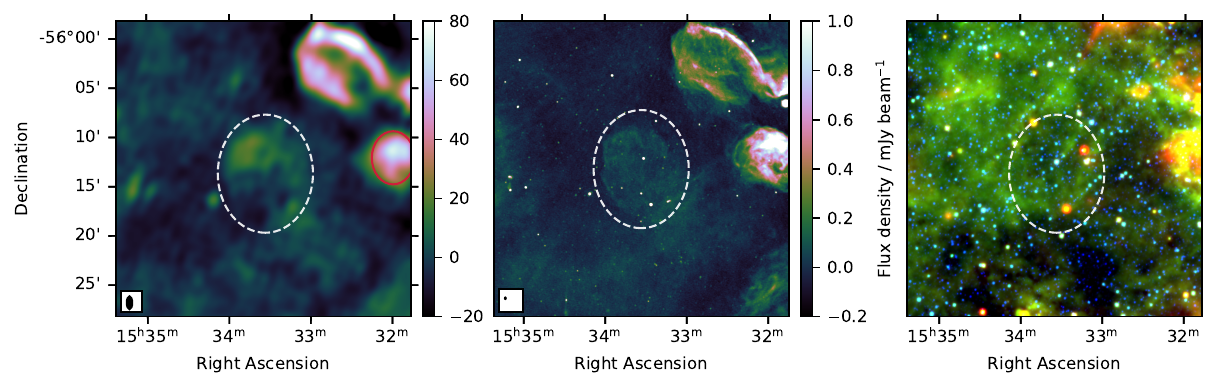}
\caption{Region surrounding \hyperref[G324.1-0.2]{G324.1$-0.2$} (class \textsc{i}) as observed by GPM at 200 MHz (left), by SMGPS at 1300 MHz (middle) and by WISE (right) at 22 $\mu$m (R), 12 $\mu$m (G), and 3.4 $\mu$m (B).}
\label{fig:G324.1-0.2}
\end{figure*}

In the left panel of \fig~\ref{fig:G324.1-0.2} is the 200\,MHz image of G324.1$-0.2$. The object appears as a filled ellipse approximately $0.14 \times 0.14$\,degrees$^2$ in size. The right panel of the same image does not show IR contamination, which increases our confidence level in a non-thermal contribution from this candidate, which has been classified as class \textsc{i}. The lower limit of the flux density is 0.24\,Jy. Recently, it has been listed in \cite{Ball2023} paper with a flux of $0.26 \pm 0.03$\,Jy at 933\,MHz; however, they do not provide a spectral index value. Using \textsc{POLYGON\_FLUX}, we then estimated a flux density of 0.20\,Jy at 1300\,MHz, which led to a spectral index of $< -0.1$. A survey with higher sensitivity and resolution is essential in making a conclusive statement on the nature of this object. 

\subsubsection{G324.1+0.0}\label{G324.1+0.0} 
G324.1+0.0 is first mentioned as a possible SNR candidate in \citet{Whiteoak1996} and described as an elongated shell with a flux density at 843\,MHz of 1.2\,Jy and a size of $14 \times 6$\,arcminutes$^2$. More recently, it has been listed in the work of \citet{Green2014} using the same telescope and observing frequency; while they measured the same size, their quoted flux densities are very different: $1.2$\,Jy and $0.3$\,Jy, respectively, even considering the error in the flux measurement, which is between 5--10\%. 

The candidate has also been detected in Johnston-Hollitt et al. (in prep.) and in the recently published work by \citet{Ball2023}, where they estimated a spectral index of $-0.3^{0.2}_{0.2}$ using flux densities at 933\,MHz and 216\,MHz. As affirmed by the authors, the uncertainty is too large to be conclusive.

\begin{figure*}
\centering
\includegraphics[width=1.0\linewidth]{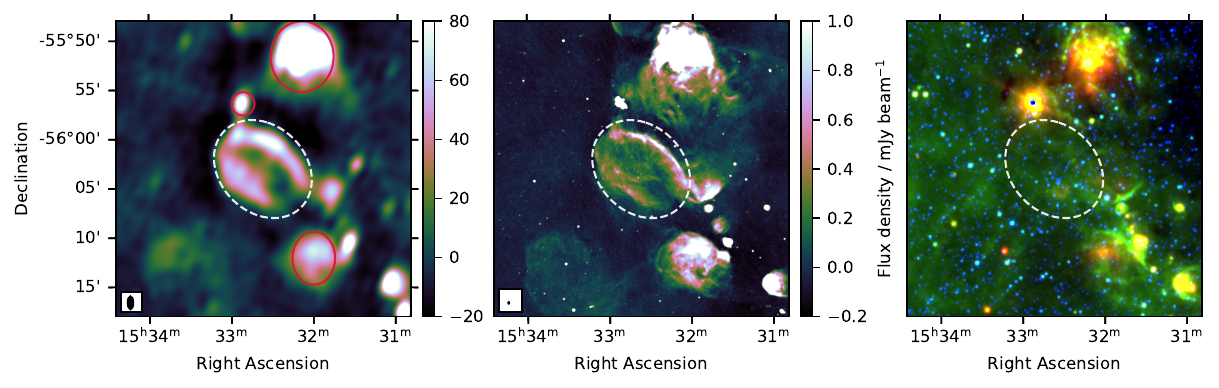}
\caption{Region surrounding \hyperref[G324.1+0.0]{G324.1+0.0} (class \textsc{i}) as observed by GPM at 200 MHz (left), by SMGPS at 1300 MHz (middle) and by WISE (right) at 22 $\mu$m (R), 12 $\mu$m (G), and 3.4 $\mu$m (B). The red ellipses surround thermal contributions in the candidate sky area.}
\label{fig:G324.1+0.0}
\end{figure*}

We estimated flux density lower limits of 1.7\,Jy at 200\,Mhz and 0.6\,Jy at 1300\,MHz. The limit for the spectral index has been calculated using the GPM limit and SMGPS data (left and middle panel of \fig~\ref{fig:G324.1+0.0}) to be greater than $< -0.6$, removing the possibility that the candidate could be dominated by thermal emission. 

%Considering a mean Galactic SNR spectral index of $-0.51$ \citep{Ranasinghe2023} and the first flux measurement mentioned from \citet{Whiteoak1996}, we estimate a value of 2.5\,Jy at 200\,MHz. The lower limit we provide is 1.7\,Jy, which is only about 68\% of the value expected from the literature. 

The candidate is located in a complex area with \textsc{hii} regions very close to it. Still, the structure itself does not seem to be affected by it, as confirmed in the right panel of \fig~\ref{fig:G324.1+0.0}. The source falls in the class~\textsc{i} of our classification scheme.

\subsubsection{G328.6+0.0}\label{G328.4-0.1} 
\begin{figure*}
\centering
\includegraphics[width=1.0\linewidth]{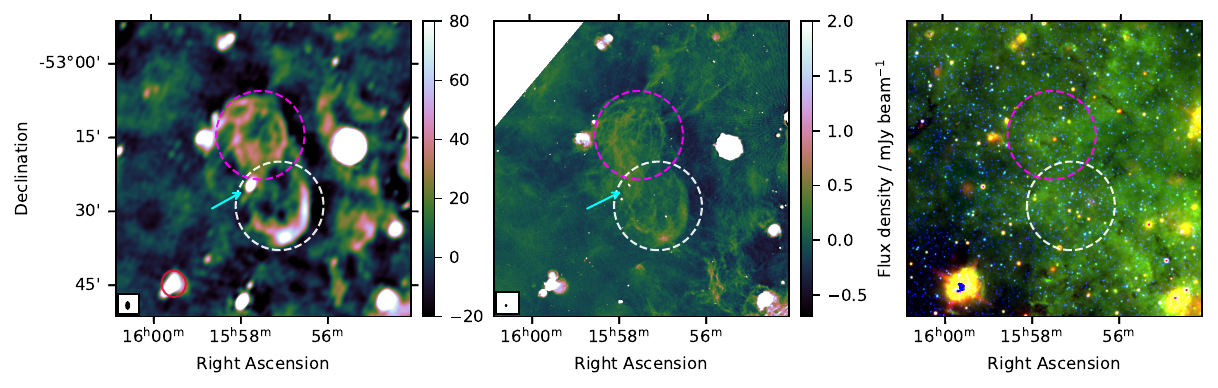}
\caption{Region surrounding \hyperref[G328.4-0.1]{G328.4$-0.1$} (class \textsc{i}) as observed by GPM at 200 MHz (left), by SMGPS at 1300 MHz (middle) and by WISE (right) at 22 $\mu$m (R), 12 $\mu$m (G), and 3.4 $\mu$m (B). The white dashed line surrounds candidate G328.4$-0.1$, and the magenta dashed line encircles the emission from a second circular structure that may be part of the same object or constitute a second SNR candidate. The red ellipse surrounds thermal contributions in the candidate sky area, while the cyan arrow points toward the position of a possible pulsar candidate.}
\label{fig:G328.4-0.1}
\end{figure*}

First noted by \citet{McClureGriffiths2001}, this SNR candidate has an ambiguous morphology that suggests the presence of two different objects not related to each other (left panel of \fig~\ref{fig:G328.4-0.1}). From a visual inspection of the GPM image, the arcs' shape seems consistent with two smaller circular structures instead of a bigger elliptical shell. The first object is centred at (15:56:58, $-53$:30:35) with the ellipse's diameters of $0.22 \times 0.26$\,degrees$^2$; while the second candidate is centred at (15:57:33, $-53$:14:42) with a radius of 0.1\,degrees. Further observations using the higher resolution SMGPS (see the middle panel in \fig~\ref{fig:G328.4-0.1}) reveal that the complex exhibits extended emission, predominantly in the east. A better sensitivity to bigger spatial scales may help clarify this object's structure. No corresponding infrared emission is detected from the sky area under examination as visible from the right panel of \fig~\ref{fig:G328.4-0.1}). The candidate has also been detected in Johnston-Hollitt et al. (in prep.) and in \cite{Ball2023}, where they calculated the flux density of the entire structure correspondent to 3.5 $\pm$ 0.9\,Jy with a spectral index of $-0.75 \pm 0.06$ which confirm that the emission is luckily non-thermal. For this reason, we classified the candidate as class~\textsc{i}, but further studies are needed to clarify the complex morphology of the object.

By comparing the images, we noticed the presence of a point-like source that does not have a thermal counterpart. We believe it to be a promising pulsar candidate; we estimated a spectral index value using GPM and SMGPS data, obtaining a value of $-1.3$, which is in accordance with the steepness of a pulsar spectrum as shown in \fig~2 of \citet{Bates2013}. Follow-up observations need to be performed to clarify the nature of the source. 

\subsubsection{G336.8$-0.6$}\label{G336.8-0.6} 
\begin{figure*}
\centering
\includegraphics[width=1.0\linewidth]{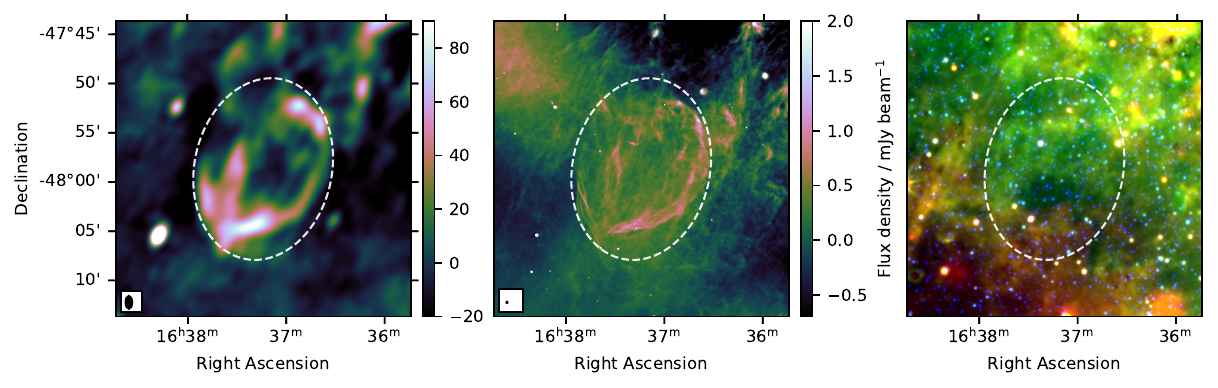}
\caption{Region surrounding \hyperref[G336.8-0.6]{G336.8$-0.6$} (class \textsc{i}) as observed by GPM at 200 MHz (left), by SMGPS at 1300 MHz (middle) and by WISE (right) at 22 $\mu$m (R), 12 $\mu$m (G), and 3.4 $\mu$m (B).}
\label{fig:G336.8-0.6}
\end{figure*}

\citet{Green2014} identified this SNR candidate at 843\,MHz as an irregular shell with bright knots at the edges and determined a flux density of 0.66 $\pm$ 0.6\,Jy with an error between 5--10\%. We observe a similar structure in the GPM data (left panel of \fig~\ref{fig:G336.8-0.6}) with the ellipse's diameters of $0.18 \times 0.28$\,degrees$^2$. A more detailed view of the candidate's filaments is given by SMGPS at 1300\,MHz and shown in the middle panel of \fig~\ref{fig:G336.8-0.6}. The RGB panel in \fig~\ref{fig:G336.8-0.6} shows clear contamination from mid-infrared emission, but no structure resembles the radio morphology. We estimated a lower limit of 1.6\,Jy at 200\,MHz and provided a limit on the spectral index correspondent to $\alpha$ $< -0.6$, which indicates that the emission is luckily non-thermal and made the candidate fall in class~\textsc{i} of our scheme. The candidate has also been detected in Johnston-Hollitt et al. (in prep.).

\subsection{New SNR candidates} \label{sub:new}
We now summarize the properties of 10~SNR candidates identified for the first time using the GPM data, listing them in \tab~\ref{tab:new}. As can be seen from \figs~\ref{fig:G16.0+0.7}-\ref{fig:G352.8-0.3}, which show the regions around each object as seen by GPM (left), RACS, EMU or SMGPS (middle), and WISE (right), the candidates are either located in a complex region of the Galactic plane, or their surface brightness is extremely low and comparable with the background itself. These characteristics explain why these SNRs were not detected in previous works. On the line of the classification scheme reported in \sect~\ref{sec:analysis}, we are confident or fairly confident that six~candidates are SNRs, attributing them a class of \textsc{i} or \textsc{ii}; the remaining four have been classified as class \textsc{iii} because the region is very confused or it shows infrared contamination such as in $G332.8-1.5$, indicating the need for further observations for a conclusive identification. 

Similarly to the candidates previously known and discussed in \sect~\ref{sub:previous}, the group of new SNR candidates is located at low latitudes; the most isolated object in this sample is $G332.8-1.5$, which resides in a very complex region of the Galaxy where thermal emission occurs, as mentioned above. They are all characterized by fairly small radii reaching a maximum of $\sim$23\,arcminutes with $G332.5-1.2$. An estimate of the lower limit for the flux densities at 200\,MHz is provided, measured using the \textsc{POLYGON\_FLUX} software and presented in \tab~\ref{tab:new}.

\begin{table*} 
\small \centering
\renewcommand{\arraystretch}{1.5}
\begin{tabular}{cccccccccccccc}\toprule
Name & RA & Dec & $l$ & $b$ & MAJ & MIN & PA & $S_{200\,MHz}$ & Morphology & Class\\ & J2000 & J2000 & $^{\circ}$ & $^{\circ}$ & $^{\circ}$ & $^{\circ}$ & $^{\circ}$ & Jy & & \\
\hline \hline   
%\hyperref[G9.9-0.2]{G9.9$-0.2$} & 18:08:37 & $-20$:24:37 & 9.99 & $-0.24$ & 0.06 & 0.13 & 60 & 2.3$\pm$0.2 & Filled & \textsc{i}\\
\hyperref[G16.0+0.7]{G16.0+0.7} & 18:17:09 & $-14$:39:13 & 16.02 & 0.74 & 0.18 & 0.18 & 0 & > 0.5 & Filled \& partial shell & \textsc{ii}\\
%G38.2+0.5 & 18:59:25 & +4:53:23 & 38.19 & 0.46 & 0.77 & 0.71 & 302 & > 3.2 & Partial shell & \textsc{iii}\\
\hyperref[G39.4-0.0]{G39.4$-0.0$} & 19:03:28 & +5:46:30 & 39.44 & $-0.03$ & 0.26 & 0.26 & 0 & > 0.6 & Shell & \textsc{iii}\\
%\hyperref[G303.4-0.7]{G303.4-0.7} & 12:55:57 & -63:36:34 & 303.44 & -0.74 & 0.16 & 0.23 & 275 & > 1.6 & Shell & \textsc{iii}\\
\hyperref[G306.4+0.1]{G306.4+0.1} & 13:21:46 & $-62$:36:36 & 306.41 & 0.06 & 0.32 & 0.32 & 0 & > 0.3 & Partial shell & \textsc{iii}\\
\hyperref[G309.2-0.1]{G309.2$-0.1$} & 13:45:44 & $-62$:19:41 & 309.19 & $-0.12$ & 0.17 & 0.23 & 340 & > 0.3 & Shell & \textsc{iii}\\
\hyperref[G332.5-1.2]{G332.5$-1.2$} & 16:21:22 & $-51$:35:48 & 332.46 & $-1.19$ & 0.39 & 0.39 & 0 & > 1.9 & Filled & \textsc{ii}\\
\hyperref[G332.8-1.5]{G332.8$-1.5$} & 16:24:25 & $-51$:32:44 & 332.83 & $-1.49$ & 0.19 & 0.19 & 0 & > 0.3 & Shell & \textsc{iii}\\
\hyperref[G333.5+0.0]{G333.5+0.0} & 16:20:36 & $-50$:01:04 & 333.49 & 0.02 & 0.18 & 0.24 & 90 & > 3.4 & Shell & \textsc{ii}\\
\hyperref[G335.7+0.9]{G335.7+0.9} & 16:26:23 & $-47$:46:59 & 335.74 & 0.92 & 0.21 & 0.21 & 0 & > 0.3 & Partial shell & \textsc{ii}\\
\hyperref[G341.4-0.2]{G341.4$-0.2$} & 16:52:49 & $-44$:16:58 & 341.41 & $-0.18$ & 0.22 & 0.22 & 0 & > 1.4 & Filled & \textsc{ii}\\
\hyperref[G352.8-0.3]{G352.8$-0.3$} & 17:28:42 & $-35$:12:11 & 352.79 & $-0.35$ & 0.20 & 0.20 & 0 & > 1.3 & Filled? & \textsc{ii}\\
\bottomrule \end{tabular} 
\caption{Properties of new SNR candidates discovered in this work, as detailed in subsection \ref{sub:new}. The table is structured as follows: \textbf{Name} derived from Galactic coordinates via \textit{lll.l} $\pm$ \textit{b.b}; \textbf{Ra, Dec} right ascension and declination in J2000 coordinates; \textbf{l, b} longitude and latitude in Galactic coordinates; \textbf{MAJ, MIN} major and minor axes of the candidate's elliptical shell in degrees; \textbf{PA} is the position angle in degree; \textbf{S$_{200}$} flux density measured at 200 MHz; \textbf{Morphology} of the shell as can be determined by visual inspection; assigned \textbf{Class} (as defined in \sect~\ref{sec:analysis}).}\label{tab:new}
\end{table*}

The following sections provide a brief description of each object in the sample.

\subsubsection{G16.0+0.7}\label{G16.0+0.7}
\begin{figure*}
\centering
\includegraphics[width=1.0\linewidth]{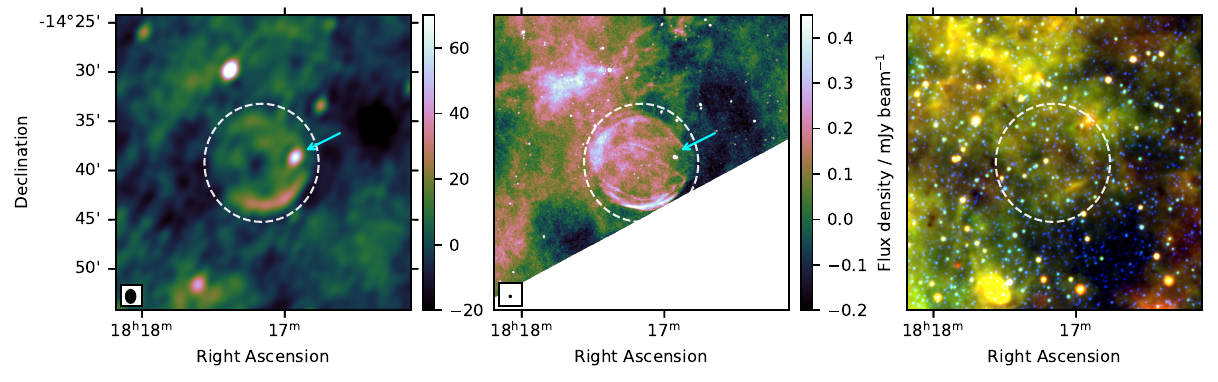}
\caption{Region surrounding \hyperref[G16.0+0.7]{G16.0+0.7} (class \textsc{ii}) as observed by GPM at 200 MHz (left), by SMGPS at 1300 MHz (middle) and by WISE (right) at 22 $\mu$m (R), 12 $\mu$m (G), and 3.4 $\mu$m (B). The cyan arrow points toward the position of a possible pulsar candidate.}
\label{fig:G16.0+0.7}
\end{figure*}

The candidate G16.0+0.7, pictured in \fig~\ref{fig:G16.0+0.7} at 200\,MHz (left panel), has an approximately circular shape with a diameter of 0.18\,degrees. The emission is particularly faint at this frequency and localized in the southern part of the shell. The SMGPS 1300\,MHz image (middle panel) shows an interesting structure in this source, resolving the southern arc into two smaller filaments. There is an absence of infrared contamination in the object's sky area (as seen in the right panel). The GPM flux density lower limit is 0.5\,Jy. We classify the remnant candidate as class~\textsc{ii}.

Inspecting the GPM and SMGPS images, we noticed the presence of a point source inside the candidate shell, which did not seem to have a thermal origin; we highlighted its location inside the shell using a cyan arrow. There is a strong possibility it could be a pulsar due to the SMGPS image revealing a potential ``blow-out'' in the right section of the shell, where the cavity could have been produced by a pulsar wind. We then performed a spectral index calculation, which gave us an estimate of $-1.2$. This value falls in the lower limit of the indices range given by \citet{Bates2013}, but to gain more confidence in the nature of the source, further observations are requested.  

%\subsubsection{G38.2+0.5}\label{G38.2+0.5}
%G38.2+0.5 has a particular morphology that recalls the shape of a triangle that, for convenience, we have reported as an ellipse with diameters of $0.77 \times 0.71$\,degrees$^2$ making it the largest object in the sample of new SNR candidates. We estimated a lower limit for the flux density of 3.2\,MHz. 

%\begin{figure*}
%\centering
%\includegraphics[width=1.0\linewidth]{SNR_G38.2+0.5.pdf}
%\caption{$2^\circ \times 2^\circ$ of the region surrounding \hyperref[G38.2+0.5]{G38.2+0.5} (class \textsc{iii}) as observed by GPM at 200 MHz (left), by SMGPS at 1300 MHz (middle) and by WISE (right) at 22 $\mu$m (R), 12 $\mu$m (G), and 3.4 $\mu$m (B).}
%\label{fig:G38.2+0.5}
%\end{figure*}

\subsubsection{G39.4$-0.0$}\label{G39.4-0.0}
\begin{figure*}
\centering
\includegraphics[width=1.0\linewidth]{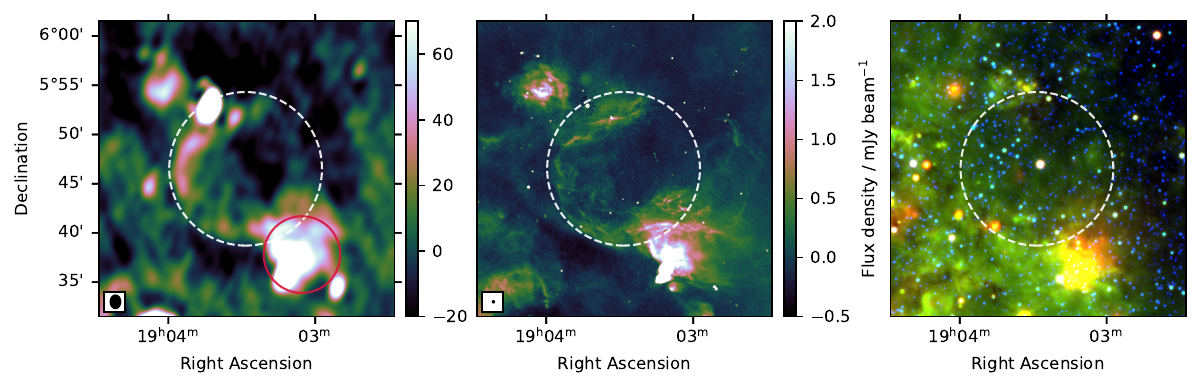}
\caption{Region surrounding \hyperref[G39.4-0.0]{G39.4$-0.0$} (class \textsc{iii}) as observed by GPM at 200 MHz (left), by SMGPS at 1300 MHz (middle) and by WISE (right) at 22 $\mu$m (R), 12 $\mu$m (G), and 3.4 $\mu$m (B). The red ellipse surrounds thermal contributions in the candidate sky area.}
\label{fig:G39.4-0.0}
\end{figure*}

This new SNR candidate has a faint circular shell with a diameter of 0.26\,degrees (as seen by GPM and SMGPS in the left and middle panel of \fig~\ref{fig:G39.4-0.0}). The right panel shows an \textsc{hii} region close to the southern east border of the shell that may contaminate the source; the RGB does not seem to have emission related to the northern part of the shell, making this object still a promising candidate. Due to the complex region where it is situated, we classified it as class~\textsc{iii}. The lower limit of the flux density is 0.6\,Jy.

%\subsubsection{G303.4-0.7}\label{G303.4-0.7}
%In the left panel of \fig~\ref{fig:G303.4-0.7} is the 200\,MHz image of G303.4-0.7. The object is a bright ellipse approximately $0.16 \times 0.23$\,degrees$^2$ in size. The right panel of the same image shows thermal contamination, particularly concentrated in the northeast region of the candidate shell. An accurate spectral index determination will be crucial in distinguishing the nature of the object in this region. Though no clear structure exists, we classified the object as part of class\,\textsc{iii}. The lower limit of the flux density is 1.6\,Jy. 

%\begin{figure*}
%\centering
%\includegraphics[width=1.0\linewidth]{SNR_G303.4-0.7.pdf}
%\caption{Region surrounding \hyperref[G303.4-0.7]{G303.4-0.7} (class \textsc{iii}) as observed by GPM at 200 MHz (left), by SMGPS at 1300 MHz (middle) and by WISE (right) at 22 $\mu$m (R), 12 $\mu$m (G), and 3.4 $\mu$m (B). The red ellipse surrounds thermal contributions in the candidate sky area.}
%\label{fig:G303.4-0.7}
%\end{figure*}

\subsubsection{G306.4+0.1}\label{G306.4+0.1}
\begin{figure*}
\centering
\includegraphics[width=1.0\linewidth]{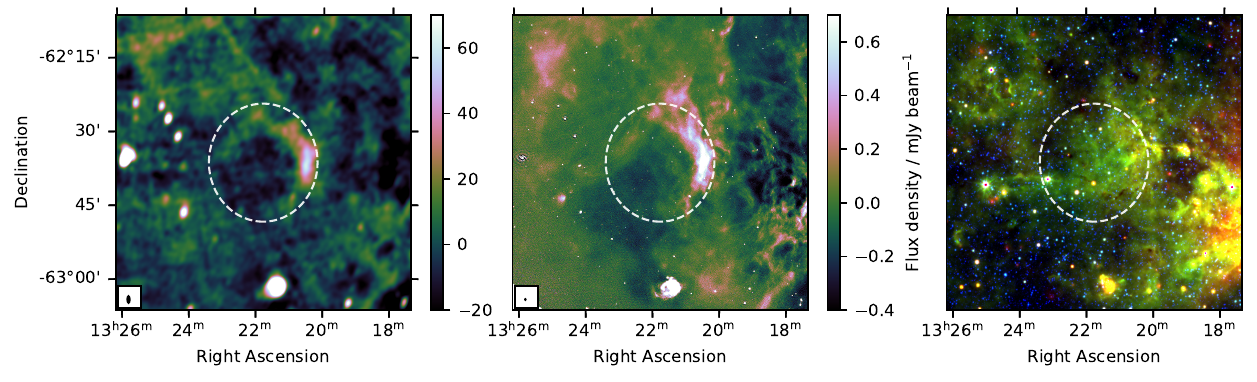}
\caption{Region surrounding \hyperref[G306.4+0.1]{G306.4+0.1} (class \textsc{iii}) as observed by GPM at 200 MHz (left), by EMU at 944 MHz (middle) and by WISE (right) at 22 $\mu$m (R), 12 $\mu$m (G), and 3.4 $\mu$m (B).}
\label{fig:G306.4+0.1}
\end{figure*}

The object has an extremely faint but almost complete circular shell of 0.32\,degrees in size, and it's shown as seen by GPM and EMU in the left and middle panel of \fig~\ref{fig:G306.4+0.1}. Most of the flux detected is concentrated in a western arc that, however, could be related to infrared emission as observed by WISE in the right panel. We still classified the object as a possible candidate of class~\textsc{iii} because the RGB does not show a complete shell as seen by the radio images. The lower limit of the flux density is 0.3\,Jy.
 
\subsubsection{G309.2-0.1}\label{G309.2$-0.1$}
\begin{figure*}
\centering
\includegraphics[width=1.0\linewidth]{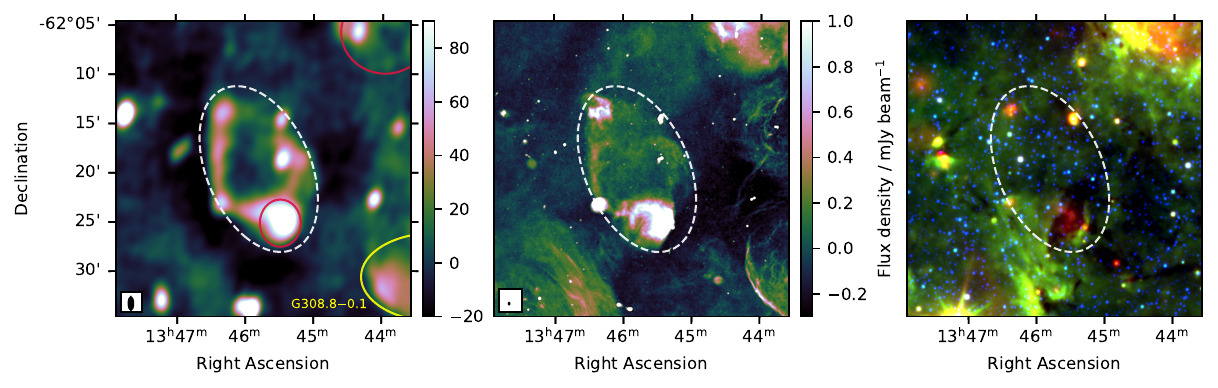}
\caption{Region surrounding \hyperref[G309.2-0.1]{G309.2$-0.1$} (class \textsc{iii}) as observed by GPM at 200 MHz (left), by SMGPS at 1300 MHz (middle) and by WISE (right) at 22 $\mu$m (R), 12 $\mu$m (G), and 3.4 $\mu$m (B). The known remnant G308.8$-0.1$ is highlighted in yellow, while the red ellipses surround thermal contributions in the candidate sky area.}
\label{fig:G309.2-0.1}
\end{figure*}

G309.2$-0.1$ appears at 200\,MHz and 1300\,MHz (left and middle panels of \fig~\ref{fig:G309.2-0.1}) as a square ring contaminated by four bright point sources around the shell and possibly an \textsc{hii} region in the southern part as confirmed by the RGB panel in \fig~\ref{fig:G309.2-0.1} where a mid-infrared emission is surrounded by an annulus ring at 12\,$\mu$m. This is probably a case where thermal and non-thermal regions are localized along the same line of sight, causing confusion and difficulty in identification. Further observations are necessary to better disentangle the emissions. The flux density has a lower limit of 0.3\,Jy and has been classified as class \textsc{iii}. The candidate has also been listed in Johnston-Hollitt et al. (in prep.).

\subsubsection{G332.5$-1.2$}\label{G332.5-1.2}
\begin{figure*}
\centering
\includegraphics[width=1.0\linewidth]{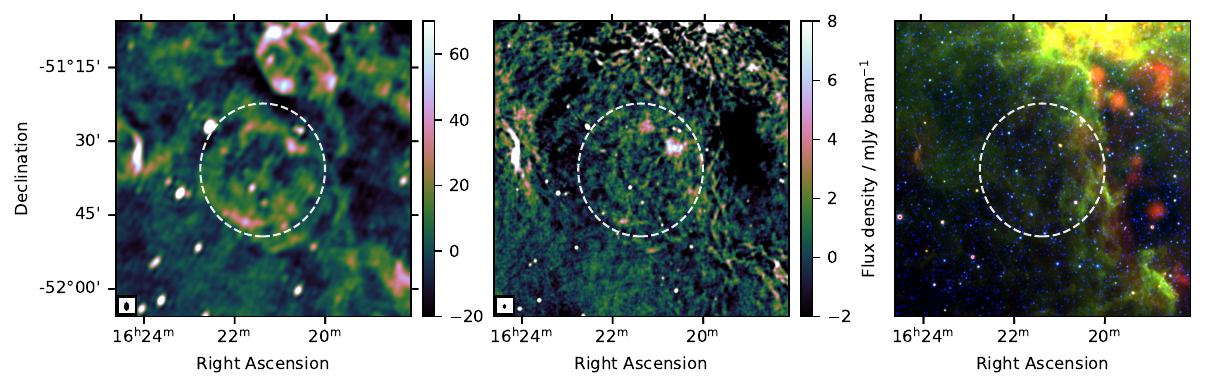}
\caption{Region surrounding \hyperref[G332.5-1.2]{G332.5$-1.2$} (class \textsc{ii}) as observed by GPM at 200 MHz (left), by RACS at 888 MHz (middle) and by WISE (right) at 22 $\mu$m (R), 12 $\mu$m (G), and 3.4 $\mu$m (B).}
\label{fig:G332.5-1.2}
\end{figure*}

\fig~\ref{fig:G332.5-1.2} shows a fairly faint and filled object of 0.39\,degrees in diameter as seen by GPM and RACS. Even if the candidate is further away from the Galactic plane, where most of the confusion is concentrated, we detect infrared emission outside the shell. This is most likely a sign of possible interaction between the suspected candidate and its surrounding environment, and therefore classify this as class~\textsc{ii}. The lower limit for the flux density is 1.9\,Jy. The candidate has also been listed in Johnston-Hollitt et al. (in prep.).

\subsubsection{G332.8$-1.5$}\label{G332.8-1.5}
\begin{figure*}
\centering
\includegraphics[width=1.0\linewidth]{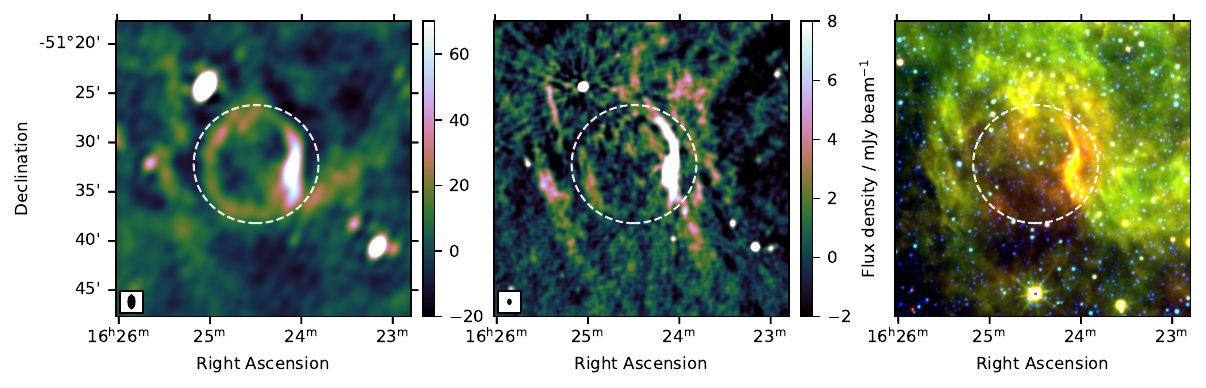}
\caption{Region surrounding \hyperref[G332.8-1.5]{G332.8$-1.5$} (class \textsc{iii}) as observed by GPM at 200 MHz (left), by RACS at 888 MHz (middle) and by WISE (right) at 22 $\mu$m (R), 12 $\mu$m (G), and 3.4 $\mu$m (B).}
\label{fig:G332.8-1.5}
\end{figure*}

G332.8$-1.5$ is the object where we had more concerns in classifying it as an SNR candidate due to the highly contaminated WISE image (see right panel of \fig~\ref{fig:G332.8-1.5}). It has a complete circular shell with a diameter of 0.19\,degrees, which makes it the second smallest object in our sample. The emission of this candidate is localized in the right arc, coincident with a bright, thermal structure. Nevertheless, G332.8$-1.5$ has still been added to the list of new SNR candidates because the remaining part of the shell does not seem to have a counterpart at infrared wavelengths (see left and middle panels) and could be another case of superimposition along the same line of sight. We classified the object as class~\textsc{iii} according to the difficulties encountered. The flux density has a lower limit of 0.3\,Jy, which does not include the bright arc coincident with thermal emission.

\subsubsection{G333.5+0.0}\label{G333.5+0.0}
\begin{figure*}
\centering
\includegraphics[width=1.0\linewidth]{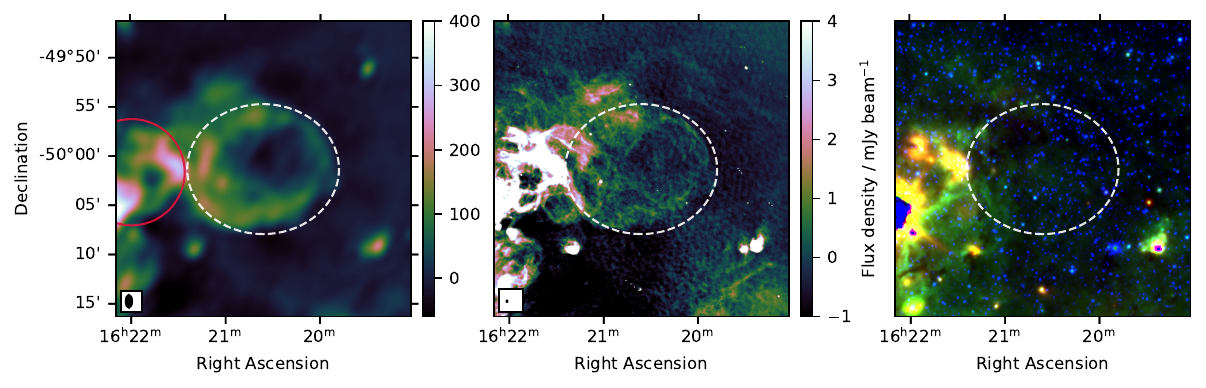}
\caption{Region surrounding \hyperref[G333.5+0.0]{G333.5+0.0} (class \textsc{ii}) as observed by GPM at 200 MHz (left), by SMGPS at 1300 MHz (middle) and by WISE (right) at 22 $\mu$m (R), 12 $\mu$m (G), and 3.4 $\mu$m (B).}
\label{fig:G333.5+0.0}
\end{figure*}

The object illustrated in \fig~\ref{fig:G333.5+0.0} by GPM data at 200\,MHz (left) and by SMGPS data at 1300\,MHz (middle) has a bright elliptical radio shell with diameters of $0.18 \times 0.24$\,degrees$^2$ and no infrared components associated with the ellipse, as can be seen in the right panel. The flux density lower limit at 200\,MHz is 3.4\,Jy. We classified the object as class \textsc{ii} for the absence of a spectral index estimate. The candidate has also been listed in Johnston-Hollitt et al. (in prep.).

\subsubsection{G335.7+0.9}\label{G335.7+0.9} %M
\begin{figure*}
\centering
\includegraphics[width=1.0\linewidth]{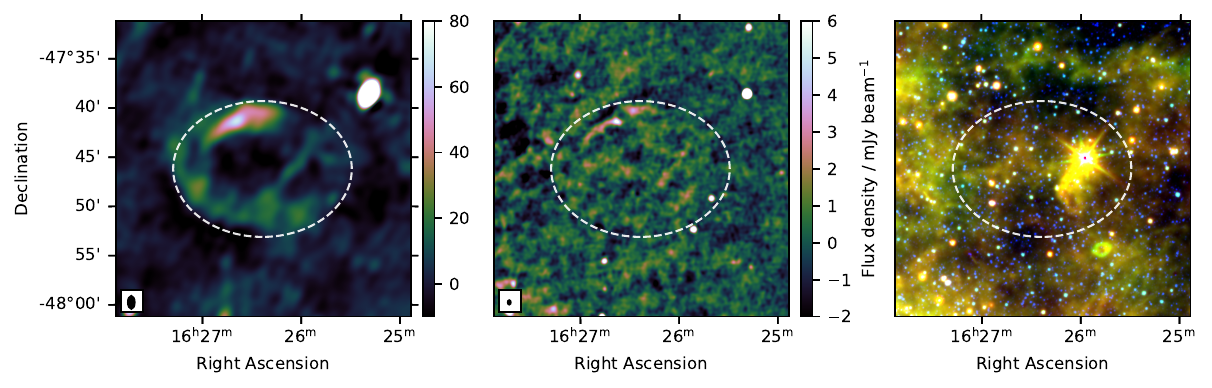}
\caption{Region surrounding \hyperref[G335.7+0.9]{G335.7+0.9} (class \textsc{ii}) as observed by GPM at 200 MHz (left), by RACS at 888 MHz (middle) and by WISE (right) at 22 $\mu$m (R), 12 $\mu$m (G), and 3.4 $\mu$m (B).}
\label{fig:G335.7+0.9}
\end{figure*}

The candidate G335.7+0.9 has a radius of 0.1\,degrees and a pretty faint surface brightness. It is hardly possible to distinguish the object from the background. It has a prominent arc in the northern region that is also resolved by the RACS survey as shown in the middle panel of \fig~\ref{fig:G335.7+0.9}. There is no infrared component to this object (right panel). The flux density lower limit is 0.3\,Jy. We classified the object as class \textsc{ii}. The candidate has also been listed in Johnston-Hollitt et al. (in prep.).

\subsubsection{G341.4$-0.2$}\label{G341.4-0.2}
\begin{figure*}
\centering
\includegraphics[width=1.0\linewidth]{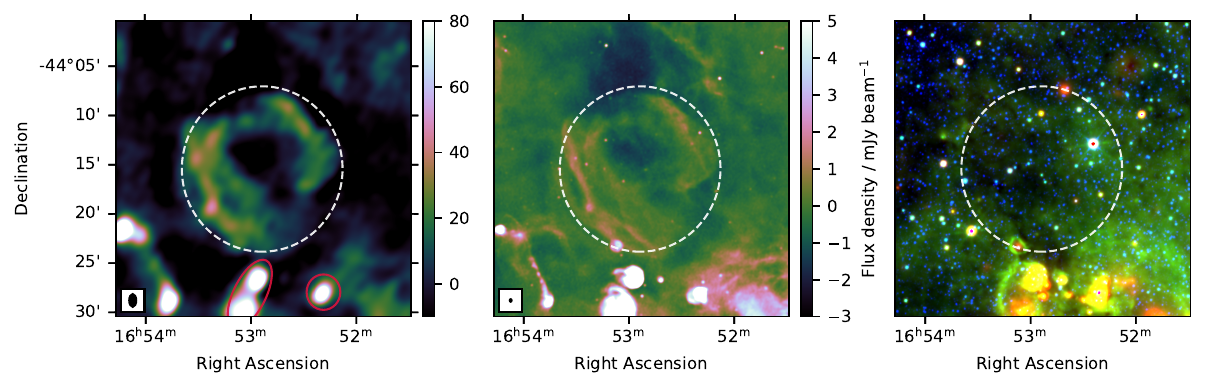}
\caption{Region surrounding \hyperref[G341.4-0.2]{G341.4$-0.2$} (class \textsc{ii}) as observed by GPM at 200 MHz (left), by EMU at 944 MHz (middle) and by WISE (right) at 22 $\mu$m (R), 12 $\mu$m (G), and 3.4 $\mu$m (B). The red ellipses surround thermal contributions in the candidate sky area.}
\label{fig:G341.4-0.2}
\end{figure*}

G341.4$-0.2$ appears as a hook in the GPM image at 200\,MHz and similarly in the EMU survey at 944\,MHz (reported in the left and middle panel of \fig~\ref{fig:G341.4-0.2}). There are no point sources or diffuse structures in the same sky area that could be the cause of contamination. The faintness of the source made us classify it only as an object of class~\textsc{ii}. The lower limit of the flux density is 1.4\,Jy. The candidate has also been listed in Johnston-Hollitt et al. (in prep.).

\subsubsection{G352.8$-0.3$}\label{G352.8-0.3}
\begin{figure*}
\centering
\includegraphics[width=1.0\linewidth]{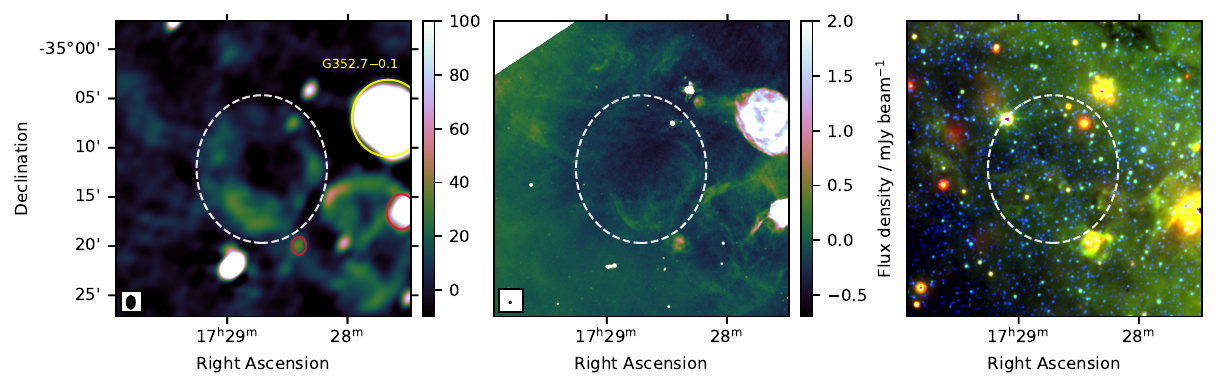}
\caption{Region surrounding \hyperref[G352.8-0.3]{G352.8$-0.3$} (class \textsc{ii}) as observed by GPM at 200 MHz (left), by SMGPS at 1300 MHz (middle) and by WISE (right) at 22 $\mu$m (R), 12 $\mu$m (G), and 3.4 $\mu$m (B). The known G352.7$-0.1$ is highlighted in yellow, while the red ellipses surround thermal contributions in the candidate sky area.}
\label{fig:G352.8-0.3}
\end{figure*}

The last candidate identified by the GPM campaign is illustrated in \fig~\ref{fig:G352.8-0.3} where it appears as a circular shell of 0.20\,degrees diameter. The middle panel of the same plot represents the candidate as seen by SMGPS at the higher frequency of 1300\,MHz, and the right panel shows that the source has no infrared components, significantly increasing the chances of this candidate being an SNR. The object has been classified as class \textsc{ii}. The flux density lower limit at 200\,MHz is 1.3\,Jy.

\section{Discussion and conclusions}\label{sec:discussion}
Our approach has enabled us to identify 20~SNR candidates in the region delimited by $285^{\circ} < l < 70^{\circ}$ using GPM data from the MWA. None of the candidates present a fully coincident infrared counterpart; six are characterized by a negative spectral index limit which suggests a dominant non-thermal emission mechanism and strengthens the source's classification as an SNR. 

\begin{figure}
\centering
\includegraphics[scale=0.35]{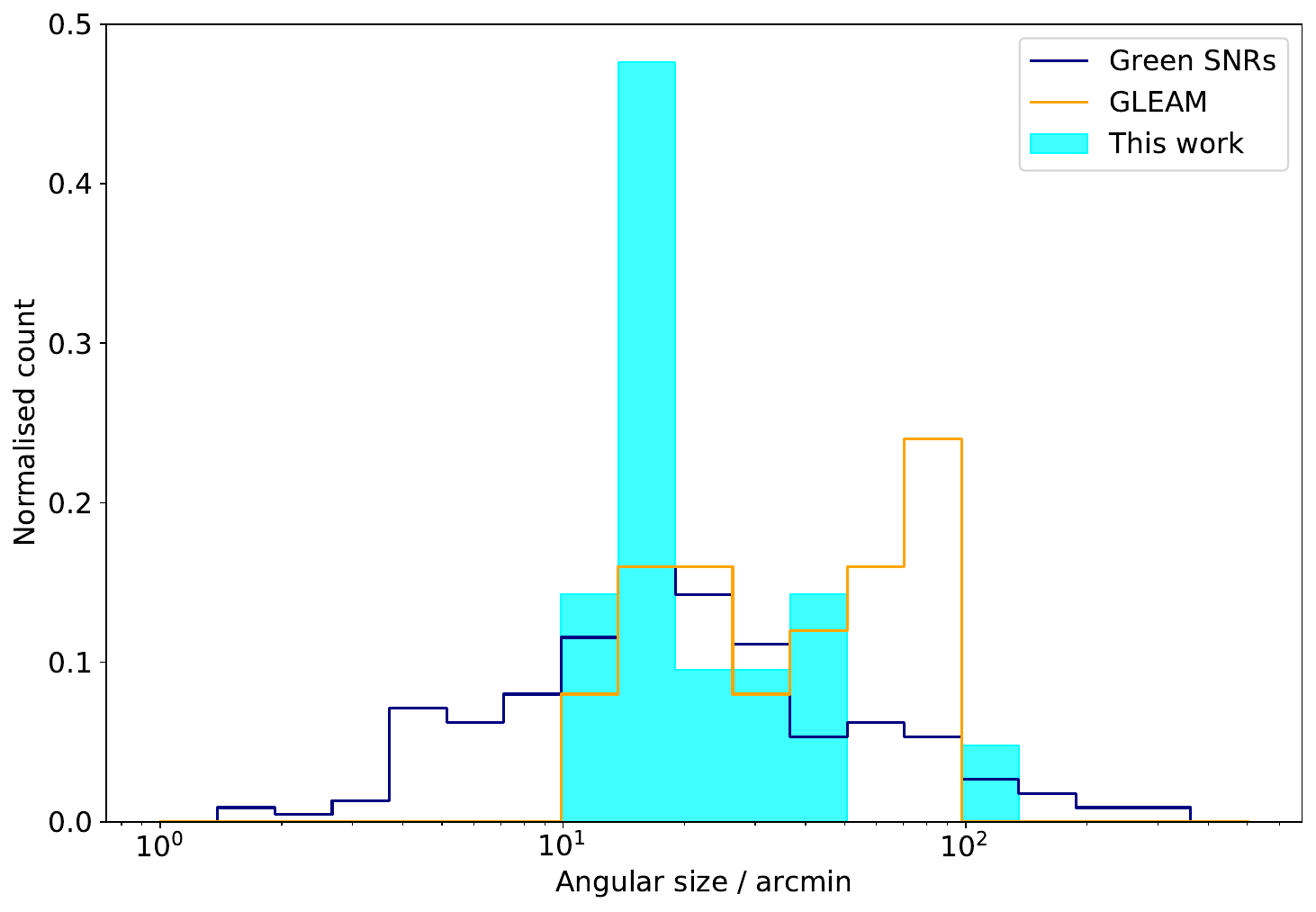}
\caption{Histogram of the candidates' angular size. It provides a comparison with known SNRs listed in the December 2022 Green catalogue and the 27 GLEAM candidates \citep{Hurley2019b}.}
\label{fig:size}
\end{figure}

Our candidates all have an angular size below 25\,arcminutes except for three candidates: G299.2$-1.5$, G310.7$-5.4$ and \\ G321.3$-3.9$. As shown in the histogram in \fig~\ref{fig:size}, the dimensions of all the elements in our sample are in agreement with the sizes expected for remnants when compared to the Green catalogue and, in general, smaller compared to the GLEAM-selected SNRs; this is expected as the MWA GLEAM survey has a resolution of ~2$'$ and sensitivity to structures of angular scales 2$'$ -- 15$^\circ$.

\begin{figure}
\centering
\includegraphics[scale=0.35]{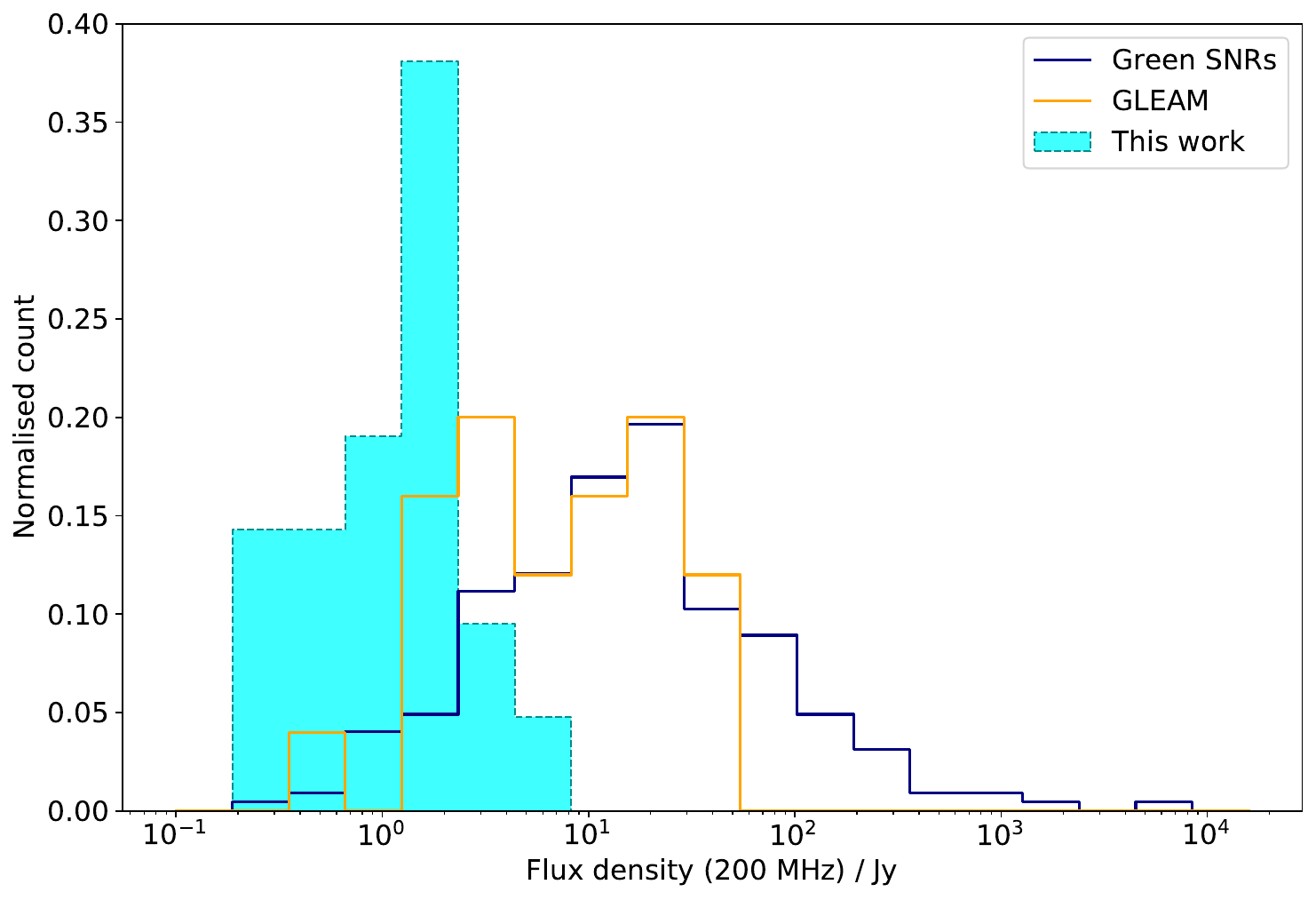}
\caption{Histogram of the candidates' flux density lower limit at 200\,MHz (dashed cyan line). It provides a comparison with known SNRs listed in the December 2022 Green catalogue (with a known spectral index value to assure an appropriate conversion of the flux to a lower frequency) and the 27 GLEAM candidates \citep{Hurley2019b}.}
\label{fig:flux}
\end{figure}

In \fig~\ref{fig:flux}, we present the flux density at 200\,MHz of the three samples of the Green catalogue, the candidates identified with the GLEAM survey and the new GPM candidates. The last mentioned sample is shown as a dashed cyan line as it corresponds to lower limits and thus cannot be directly compared with the known SNRs and the GLEAM candidates, which provide exact values. Even if we are missing 50\% of the flux of our new candidates, they reside in the lower range of flux densities of the objects reported by Green.
The faintness of the SNR candidates explains why these objects were not confirmed in the Green catalogue and highlights how a study with better sensitivity to all angular scales will increase our possibility of finding the missing population.

\begin{figure*}
\centering
\subfloat[]{\includegraphics[scale=0.39]{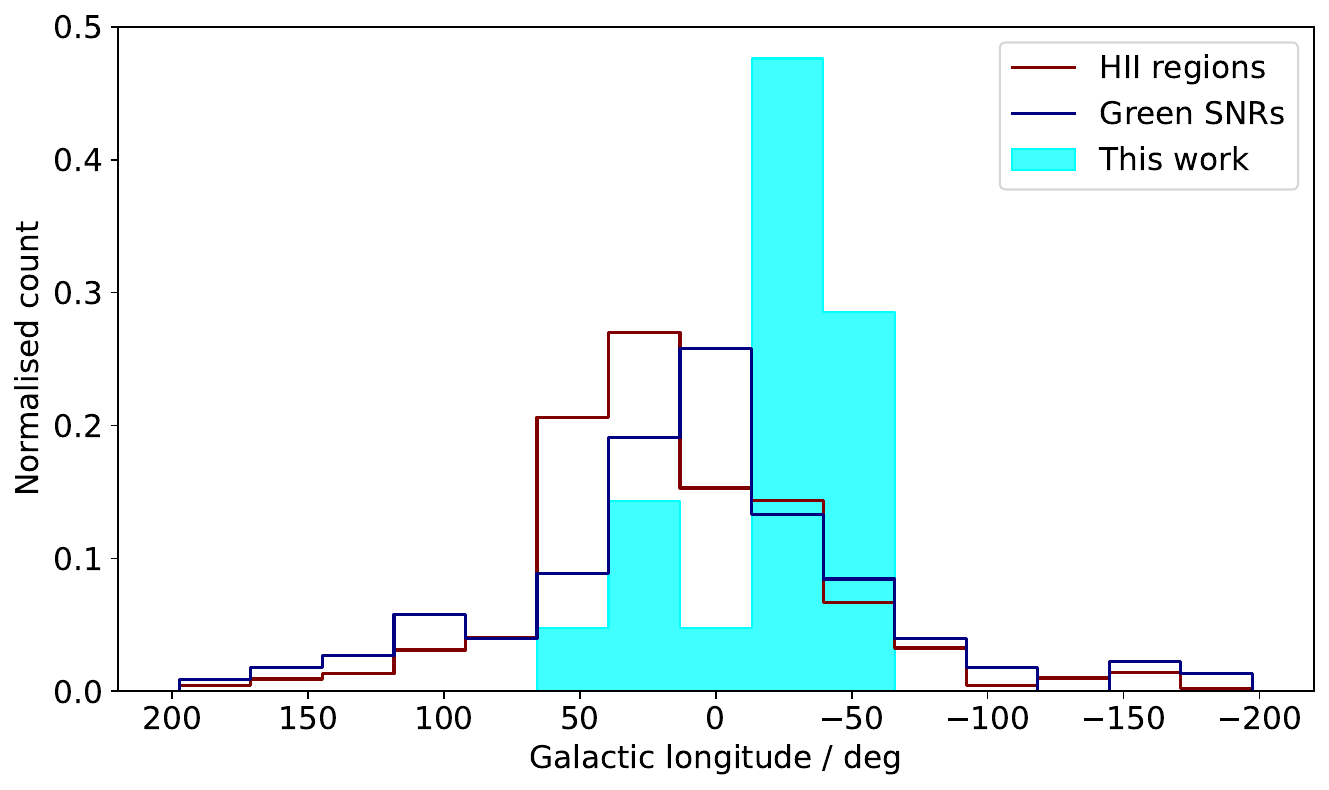}} \quad
\subfloat[]{\includegraphics[scale=0.39]{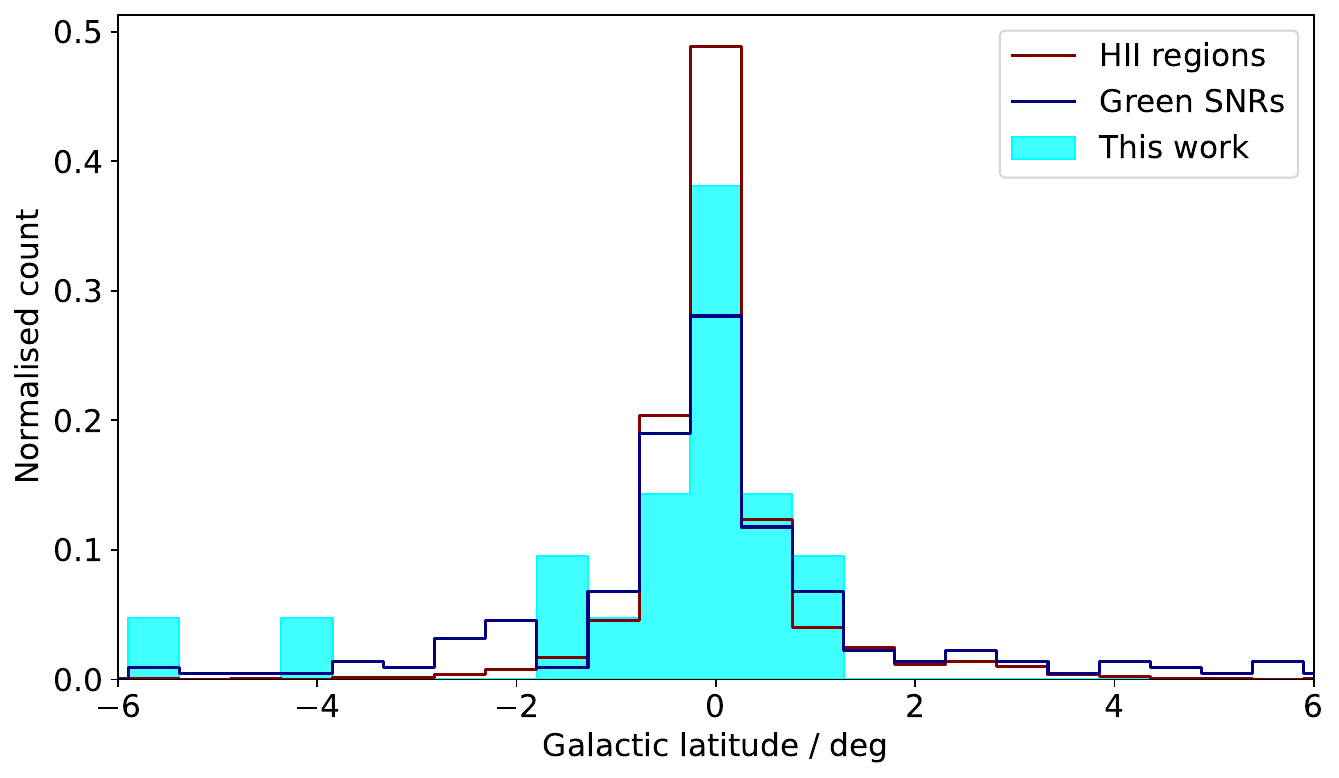}}
\caption{Distribution of the SNR candidates identified in this work as a function of the longitude (panel \textsc{a}) and latitude (panel \textsc{b}). The distribution along the Galactic plane of the known SNRs (as per the December 2022 version of Green's catalogue) is represented in blue, while the red histogram illustrates the distribution of \textsc{hii} regions from the V2.2 WISE catalogue. The counts have been normalized to form a probability density.}
\label{fig:distribution}
\end{figure*}

We now compare the Galactic longitude and latitude of our SNR candidates with the large Green SNR sample and \textsc{hii} regions in order to examine their spatial distribution and investigate potential correlations between their locations. This comparison may also help us identify any systematic biases or selection effects in our sample. As shown in \fig~\ref{fig:distribution}, where the longitude (left panel) and latitude (right panel) locations are represented, the detected sample follows a similar distribution to the Green SNRs. The V2.2 WISE catalogue of \textsc{hii} regions along the Galactic plane has been downloaded~\footnote{http://astro.phys.wvu.edu/wise/} and included in the analysis; albeit \textsc{hii} regions follow the same latitude trend limited within a modulus of five degrees, the longitude distribution has a similar shape to SNRs except around $l=\ang{45}$ where they show a peak indicating the youths of the W43 star formation region. On the other hand, close to $b=\ang{0}$, the candidate histogram does not show a change in the inclination when compared to the Green catalogue and the \textsc{hii} region distribution. It also highlights how our work has succeeded in detecting a higher fraction of candidates in this area compared to the SNR distribution. Nevertheless, we are still missing many candidates, assuming the distribution resembles that of \textsc{hii} regions where almost 50\% of the sources are localized.  

It is worth mentioning the work done by \citet{Dokara2021}, in which the authors identified 157~SNR candidates using the GLOSTAR survey of the Galactic plane. They provide a similar latitude histogram showing a small discrepancy between the distribution of the candidates detected compared to the known SNRs and \textsc{hii} regions. The two known samples present a histogram peak centre in $b=\ang{0}$, with a slightly higher count for negative latitudes, in opposition to their sample of candidates that is shifted toward positive values. Instead, the candidates identified in this work seem to follow the distribution of the Green catalogue objects and the thermal sources, showing a higher number of sources for $b < 0^\circ$. However, our sample is not statistically significant, and we can not formulate a conclusive statement. It is, therefore, still unclear what can cause the bias \citet{Dokara2021} are facing in their analysis and whether it is expected to occur; increasing the overall population of SNR in the Galaxy will help to clarify whether it is a selection or a real effect.

If the candidates are confirmed, compared to the Green catalogue, this work increases the remnant population by 9\% in the same region of the sky over $ 285^{\circ} < l < 70^{\circ}$ and $|b| < 16^{\circ}$, and the total number of SNRs by 7\%. Of the 20 candidates in our sample, it has only been possible to provide a spectral index estimate for six; in the rest of the cases, the radio surveys covering the candidate's locations do not fully measure structures of their size. We are currently working on a joint deconvolution of data collected by the GLEAM survey and the GLEAM-X \citep[eXtended; ][]{Hurley2022,Ross2024} survey to take advantage of the high resolution and sensitivity to all spatial scales ($45^{''} - 15^{\circ}$ and frequency range 72--231\,MHz) that these two data sets can provide. Using these data, we will be able to measure the spectral indices of these and other candidates, as well as detect new candidates, further decreasing the discrepancy between theory and observations. By combining the results with the recently unveiled SMGPS survey and the gradually emerging EMU survey data, which have already helped in the confirmation of some of the candidates presented in this work and will definitely consent to further detections, we will be able to provide flux density measurements spanning 72\,MHz to 1.4\,GHz. This comprehensive dataset will enable an excellent sampling of SNRs that will only be superseded by the SKA.

Moreover, multi-wavelength studies are essential in understanding the phenomena we observe in the sky, and the advantages are multiple. Every wavelength provides different information on the sources under examination. By taking a multifrequency approach, we obtain more precise estimates of the physical parameters, giving a more complete view of the mechanisms and underlying physics at work. Future work will make use of the fourth data release of the Mopra Southern Galactic Plane carbon monoxide (CO) survey \citep{Cubuk2023} to provide useful information on the composition of the environment surrounding the candidates we have described in this paper, highlighting the possible presence of a molecular cloud in the vicinity and so probing its interactions with the shock fronts, and providing distance estimates.

\begin{acknowledgement}
We thank Roland Kothes, Miroslav Filipovic and Melanie Johnston-Hollitt for the judicious feedback on this paper.
NHW is supported by an Australian Research Council Future Fellowship (project number FT190100231) funded by the Australian Government.
This publication makes use of data products from the Wide-field Infrared Survey Explorer, which is a joint project of the University of California, Los Angeles, and the Jet Propulsion Laboratory/California Institute of Technology, and NEOWISE, which is a project of the Jet Propulsion Laboratory/California Institute of Technology. WISE and NEOWISE are funded by the National Aeronautics and Space Administration.

This scientific work uses data obtained from Inyarrimanha Ilgari Bundara / the Murchison Radio-astronomy Observatory. We acknowledge the Wajarri Yamaji People as the Traditional Owners and native title holders of the Observatory site. CSIRO’s ASKAP radio telescope is part of the Australia Telescope National Facility (https://ror.org/05qajvd42). Operation of ASKAP is funded by the Australian Government with support from the National Collaborative Research Infrastructure Strategy. ASKAP uses the resources of the Pawsey Supercomputing Research Centre. Establishment of ASKAP, Inyarrimanha Ilgari Bundara, the CSIRO Murchison Radio-astronomy Observatory and the Pawsey Supercomputing Research Centre are initiatives of the Australian Government, with support from the Government of Western Australia and the Science and Industry Endowment Fund. This paper includes archived data obtained through the CSIRO ASKAP Science Data Archive, CASDA (https://data.csiro.au).
\end{acknowledgement}

\printendnotes
\printbibliography

@ARTICLE{Ball2023,
author = {{Ball}, Brianna D. and {Kothes}, Roland and {Rosolowsky}, Erik and {West}, Jennifer and {Becker}, Werner and {Filipovi{\'c}}, Miroslav D. and {Gaensler}, B.~M. and {Hopkins}, Andrew M. and {Koribalski}, B{\"a}rbel and {Landecker}, Tom and {Leahy}, Denis and {Marvil}, Joshua and {Sun}, Xiaohui and {Bufano}, Filomena and {Carretti}, Ettore and {Ingallinera}, Adriano and {Van Eck}, Cameron L. and {Willis}, Tony},
title = "{A catalogue of radio supernova remnants and candidate supernova remnants in the EMU/POSSUM Galactic pilot field}",
journal = {\mnras},
year = 2023,
month = sep,
volume = {524},
number = {1},
pages = {1396-1421},
doi = {10.1093/mnras/stad1953},
archivePrefix = {arXiv},
eprint = {2307.01948},
primaryClass = {astro-ph.GA},
adsurl = {https://ui.adsabs.harvard.edu/abs/2023MNRAS.524.1396B}
}

@ARTICLE{Bates2013,
author = {{Bates}, S.~D. and {Lorimer}, D.~R. and {Verbiest}, J.~P.~W.},
title = "{The pulsar spectral index distribution}",
journal = {\mnras},
year = 2013,
month = may,
volume = {431},
number = {2},
pages = {1352-1358},
doi = {10.1093/mnras/stt257},
archivePrefix = {arXiv},
eprint = {1302.2053},
primaryClass = {astro-ph.SR},
adsurl = {https://ui.adsabs.harvard.edu/abs/2013MNRAS.431.1352B}
}

@INPROCEEDINGS{Bertin2002,
author = {{Bertin}, Emmanuel and {Mellier}, Yannick and {Radovich}, Mario and {Missonnier}, Gilles and {Didelon}, Pierre and {Morin}, Bertrand},
title = "{The TERAPIX Pipeline}",
booktitle = {Astronomical Data Analysis Software and Systems XI},
year = 2002,
editor = {{Bohlender}, David A. and {Durand}, Daniel and {Handley}, Thomas H.},
series = {Astronomical Society of the Pacific Conference Series},
volume = {281},
month = jan,
pages = {228},
adsurl = {https://ui.adsabs.harvard.edu/abs/2002ASPC..281..228B}
}

@ARTICLE{Brogan2006,
author = {{Brogan}, C.~L. and {Gelfand}, J.~D. and {Gaensler}, B.~M. and {Kassim}, N.~E. and {Lazio}, T.~J.~W.},
title = "{Discovery of 35 New Supernova Remnants in the Inner Galaxy}",
journal = {\apjl},
year = 2006,
month = mar,
volume = {639},
number = {1},
pages = {L25-L29},
doi = {10.1086/501500},
archivePrefix = {arXiv},
eprint = {astro-ph/0601451},
primaryClass = {astro-ph},
adsurl = {https://ui.adsabs.harvard.edu/abs/2006ApJ...639L..25B}
}

@ARTICLE{Brown1952,
author = {{Hanbury Brown}, R. and {Hazard}, C.},
title = "{Radio-Frequency Radiation from Tycho Brahe's Supernova (A.D. 1572)}",
journal = {\nat},
year = 1952,
month = aug,
volume = {170},
number = {4322},
pages = {364-365},
doi = {10.1038/170364a0},
adsurl = {https://ui.adsabs.harvard.edu/abs/1952Natur.170..364H}
}

@ARTICLE{Calabretta2014,
author = {{Calabretta}, Mark R. and {Staveley-Smith}, Lister and {Barnes}, David G.},
title = "{A New 1.4 GHz Radio Continuum Map of the Sky South of Declination +25{\textdegree}}",
journal = {\pasa},
year = 2014,
month = jan,
volume = {31},
eid = {e007},
pages = {e007},
doi = {10.1017/pasa.2013.36},
archivePrefix = {arXiv},
eprint = {1310.2414},
primaryClass = {astro-ph.IM},
adsurl = {https://ui.adsabs.harvard.edu/abs/2014PASA...31....7C}
}

@ARTICLE{Carretti2019,
author = {{Carretti}, E. and {Haverkorn}, M. and {Staveley-Smith}, L. and {Bernardi}, G. and {Gaensler}, B.~M. and {Kesteven}, M.~J. and {Poppi}, S. and {Brown}, S. and {Crocker}, R.~M. and {Purcell}, C. and {Schnitzeler}, D.~H.~F.~M. and {Sun}, X.},
title = "{S-band Polarization All-Sky Survey (S-PASS): survey description and maps}",
journal = {\mnras},
year = 2019,
month = oct,
volume = {489},
number = {2},
pages = {2330-2354},
doi = {10.1093/mnras/stz806},
archivePrefix = {arXiv},
eprint = {1903.09420},
primaryClass = {astro-ph.GA},
adsurl = {https://ui.adsabs.harvard.edu/abs/2019MNRAS.489.2330C}
}

@BOOK{Clark1977,
author = {{Clark}, David H. and {Stephenson}, Francis Richard},
title = "{The historical supernovae}",
year = 1977,
adsurl = {https://ui.adsabs.harvard.edu/abs/1977hisu.book.....C}
}

@BOOK{Condon2016,
author = {{Condon}, James J. and {Ransom}, Scott M.},
title = "{Essential Radio Astronomy}",
year = 2016,
adsurl = {https://ui.adsabs.harvard.edu/abs/2016era..book.....C}
}

@ARTICLE{Cubuk2023,
author = {{Cubuk}, K.~O. and {Burton}, M.~G. and {Braiding}, C. and {Wong}, G.~F. and {Rowell}, G. and {Maxted}, N.~I. and {Eden}, D. and {Alsaberi}, R.~Z.~E. and {Blackwell}, R. and {Enokiya}, R. and {Feijen}, K. and {Filipovi{\'c}}, M.~D. and {Freeman}, M.~S.~R. and {Fujita}, S. and {Ghavam}, M. and {Gunay}, B. and {Indermuehle}, B. and {Hayashi}, K. and {Kohno}, M. and {Nagaya}, T. and {Nishimura}, A. and {Okawa}, K. and {Rebolledo}, D. and {Romano}, D. and {Sano}, H. and {Snoswell}, C. and {Tothill}, N.~F.~H. and {Tsuge}, K. and {Voisin}, F. and {Yamane}, Y. and {Yoshiike}, S.},
title = "{The Mopra Southern Galactic Plane CO Survey - data release 4- complete survey}",
journal = {\pasa},
year = 2023,
month = sep,
volume = {40},
eid = {e047},
pages = {e047},
doi = {10.1017/pasa.2023.44},
adsurl = {https://ui.adsabs.harvard.edu/abs/2023PASA...40...47C}
}

@ARTICLE{Dewdney2009,
author = {{Dewdney}, P.~E. and {Hall}, P.~J. and {Schilizzi}, R.~T. and {Lazio}, T.~J.~L.~W.},
title = "{The Square Kilometre Array}",
journal = {IEEE Proceedings},
year = 2009,
month = aug,
volume = {97},
number = {8},
pages = {1482-1496},
doi = {10.1109/JPROC.2009.2021005},
adsurl = {https://ui.adsabs.harvard.edu/abs/2009IEEEP..97.1482D}
}

@ARTICLE{Dokara2021,
author = {{Dokara}, R. and {Brunthaler}, A. and {Menten}, K.~M. and {Dzib}, S.~A. and {Reich}, W. and {Cotton}, W.~D. and {Anderson}, L.~D. and {Chen}, C. -H.~R. and {Gong}, Y. and {Medina}, S. -N.~X. and {Ortiz-Le{\'o}n}, G.~N. and {Rugel}, M. and {Urquhart}, J.~S. and {Wyrowski}, F. and {Yang}, A.~Y. and {Beuther}, H. and {Billington}, S.~J. and {Csengeri}, T. and {Carrasco-Gonz{\'a}lez}, C. and {Roy}, N.},
title = "{A global view on star formation: The GLOSTAR Galactic plane survey. II. Supernova remnants in the first quadrant of the Milky Way}",
journal = {\aap},
year = 2021,
month = jul,
volume = {651},
eid = {A86},
pages = {A86},
doi = {10.1051/0004-6361/202039873},
archivePrefix = {arXiv},
eprint = {2103.06267},
primaryClass = {astro-ph.GA},
adsurl = {https://ui.adsabs.harvard.edu/abs/2021A&A...651A..86D}
}

@ARTICLE{Dubner2015,
author = {{Dubner}, Gloria and {Giacani}, Elsa},
title = "{Radio emission from supernova remnants}",
journal = {\aapr},
year = 2015,
month = sep,
volume = {23},
eid = {3},
pages = {3},
doi = {10.1007/s00159-015-0083-5},
archivePrefix = {arXiv},
eprint = {1508.07294},
primaryClass = {astro-ph.HE},
adsurl = {https://ui.adsabs.harvard.edu/abs/2015A&ARv..23....3D}
}

@ARTICLE{Duncan1997,
author = {{Duncan}, A.~R. and {Stewart}, R.~T. and {Haynes}, R.~F. and {Jones}, K.~L.},
title = "{Supernova remnant candidates from the Parkes 2.4-GHz survey}",
journal = {\mnras},
year = 1997,
month = jun,
volume = {287},
number = {4},
pages = {722-738},
doi = {10.1093/mnras/287.4.722},
adsurl = {https://ui.adsabs.harvard.edu/abs/1997MNRAS.287..722D}
}

@ARTICLE{Duyvendak1942,
author = {{Duyvendak}, J.~J.~L.},
title = "{Further Data Bearing on the Identification of the Crab Nebula with the Supernova of 1054 A.D. Part I. The Ancient Oriental Chronicles}",
journal = {\pasp},
year = 1942,
month = apr,
volume = {54},
number = {318},
pages = {91-94},
doi = {10.1086/125409},
adsurl = {https://ui.adsabs.harvard.edu/abs/1942PASP...54...91D}
}

@ARTICLE{Ferrand2012,
author = {{Ferrand}, Gilles and {Safi-Harb}, Samar},
title = "{A census of high-energy observations of Galactic supernova remnants}",
journal = {Advances in Space Research},
year = 2012,
month = may,
volume = {49},
number = {9},
pages = {1313-1319},
doi = {10.1016/j.asr.2012.02.004},
archivePrefix = {arXiv},
eprint = {1202.0245},
primaryClass = {astro-ph.HE},
adsurl = {https://ui.adsabs.harvard.edu/abs/2012AdSpR..49.1313F}
}

@ARTICLE{Frail1994,
author = {{Frail}, D.~A. and {Goss}, W.~M. and {Whiteoak}, J.~B.~Z.},
title = "{The Radio Lifetime of Supernova Remnants and the Distribution of Pulsar Velocities at Birth}",
journal = {\apj},
year = 1994,
month = dec,
volume = {437},
pages = {781},
doi = {10.1086/175038},
archivePrefix = {arXiv},
eprint = {astro-ph/9407031},
primaryClass = {astro-ph},
adsurl = {https://ui.adsabs.harvard.edu/abs/1994ApJ...437..781F}
}

@INPROCEEDINGS{Gaensler2010,
author = {{Gaensler}, Bryan M. and {Landecker}, T.~L. and {Taylor}, A.~R. and {POSSUM Collaboration}},
title = "{Survey Science with ASKAP: Polarization Sky Survey of the Universe's Magnetism (POSSUM)}",
booktitle = {American Astronomical Society Meeting Abstracts \#215},
year = 2010,
series = {American Astronomical Society Meeting Abstracts},
volume = {215},
month = jan,
eid = {470.13},
pages = {470.13},
adsurl = {https://ui.adsabs.harvard.edu/abs/2010AAS...21547013G}
}

@ARTICLE{Green1999,
author = {{Green}, A.~J. and {Cram}, L.~E. and {Large}, M.~I. and {Ye}, Taisheng},
title = "{The Molonglo Galactic Plane Survey. I. Overview and Images}",
journal = {\apjs},
year = 1999,
month = may,
volume = {122},
number = {1},
pages = {207-219},
doi = {10.1086/313208},
archivePrefix = {arXiv},
eprint = {astro-ph/9810385},
primaryClass = {astro-ph},
adsurl = {https://ui.adsabs.harvard.edu/abs/1999ApJS..122..207G}
}

@ARTICLE{Green2014,
author = {{Green}, A.~J. and {Reeves}, S.~N. and {Murphy}, T.},
title = "{The Second Epoch Molonglo Galactic Plane Survey: Images and Candidate Supernova Remnants}",
journal = {\pasa},
year = 2014,
month = nov,
volume = {31},
eid = {e042},
pages = {e042},
doi = {10.1017/pasa.2014.37},
archivePrefix = {arXiv},
eprint = {1410.8247},
primaryClass = {astro-ph.GA},
adsurl = {https://ui.adsabs.harvard.edu/abs/2014PASA...31...42G}
}

@ARTICLE{Green2019,
author = {{Green}, D.~A.},
title = "{A revised catalogue of 294 Galactic supernova remnants}",
journal = {Journal of Astrophysics and Astronomy},
year = 2019,
month = aug,
volume = {40},
number = {4},
eid = {36},
pages = {36},
doi = {10.1007/s12036-019-9601-6},
archivePrefix = {arXiv},
eprint = {1907.02638},
primaryClass = {astro-ph.GA},
adsurl = {https://ui.adsabs.harvard.edu/abs/2019JApA...40...36G}
}

@ARTICLE{Goedhart2024,
author = {{Goedhart}, S. and {Cotton}, W.~D. and {Camilo}, F. and {Thompson}, M.~A. and {Umana}, G. and {Bietenholz}, M. and {Woudt}, P.~A. and {Anderson}, L.~D. and {Bordiu}, C. and {Buckley}, D.~A.~H. and {Buemi}, C.~S. and {Bufano}, F. and {Cavallaro}, F. and {Chen}, H. and {Chibueze}, J.~O. and {Egbo}, D. and {Frank}, B.~S. and {Hoare}, M.~G. and {Ingallinera}, A. and {Irabor}, T. and {Kraan-Korteweg}, R.~C. and {Kurapati}, S. and {Leto}, P. and {Loru}, S. and {Mutale}, M. and {Obonyo}, W.~O. and {Plavin}, A. and {Rajohnson}, S.~H.~A. and {Rigby}, A. and {Riggi}, S. and {Seidu}, M. and {Serra}, P. and {Smart}, B.~M. and {Stappers}, B.~W. and {Steyn}, N. and {Surnis}, M. and {Trigilio}, C. and {Williams}, G.~M. and {Abbott}, T.~D. and {Adam}, R.~M. and {Asad}, K.~M.~B. and {Baloyi}, T. and {Bauermeister}, E.~F. and {Bennet}, T.~G.~H. and {Bester}, H. and {Botha}, A.~G. and {Brederode}, L.~R.~S. and {Buchner}, S. and {Burger}, J.~P. and {Cheetham}, T. and {Cloete}, K. and {de Villiers}, M.~S. and {de Villiers}, D.~I.~L. and {du Toit}, L.~J. and {Esterhuyse}, S.~W.~P. and {Fanaroff}, B.~L. and {Fourie}, D.~J. and {Gamatham}, R.~R.~G. and {Gatsi}, T.~G. and {Geyer}, M. and {Gouws}, M. and {Gumede}, S.~C. and {Heywood}, I. and {Hokwana}, A. and {Hoosen}, S.~W. and {Horn}, D.~M. and {Horrell}, L.~M.~G. and {Hugo}, B.~V. and {Isaacson}, A.~I. and {J{\'o}zsa}, G.~I.~G. and {Jonas}, J.~L. and {Jordaan}, J.~D.~B.~L. and {Joubert}, A.~F. and {Julie}, R.~P.~M. and {Kapp}, F.~B. and {Kriek}, N. and {Kriel}, H. and {Krishnan}, V.~K. and {Kusel}, T.~W. and {Legodi}, L.~S. and {Lehmensiek}, R. and {Lord}, R.~T. and {Macfarlane}, P.~S. and {Magnus}, L.~G. and {Magozore}, C. and {Main}, J.~P.~L. and {Malan}, J.~A. and {Manley}, J.~R. and {Marais}, S.~J. and {Maree}, M.~D.~J. and {Martens}, A. and {Maruping}, P. and {McAlpine}, K. and {Merry}, B.~C. and {Mgodeli}, M. and {Millenaar}, R.~P. and {Mokone}, O.~J. and {Monama}, T.~E. and {New}, W.~S. and {Ngcebetsha}, B. and {Ngoasheng}, K.~J. and {Nicolson}, G.~D. and {Ockards}, M.~T. and {Oozeer}, N. and {Passmoor}, S.~S. and {Patel}, A.~A. and {Peens-Hough}, A. and {Perkins}, S.~J. and {Ramaila}, A.~J.~T. and {Ratcliffe}, S.~M. and {Renil}, R. and {Richter}, L.~L. and {Salie}, S. and {Sambu}, N. and {Schollar}, C.~T.~G. and {Schwardt}, L.~C. and {Schwartz}, R.~L. and {Serylak}, M. and {Siebrits}, R. and {Sirothia}, S.~K. and {Slabber}, M.~J. and {Smirnov}, O.~M. and {Tiplady}, A.~J. and {van Balla}, T.~J. and {van der Byl}, A. and {Van Tonder}, V. and {Venter}, A.~J. and {Venter}, M. and {Welz}, M.~G. and {Williams}, L.~P.},
title = "{The SARAO MeerKAT 1.3 GHz Galactic Plane Survey}",
journal = {\mnras},
year = 2024,
month = jun,
volume = {531},
number = {1},
pages = {649-681},
doi = {10.1093/mnras/stae1166},
archivePrefix = {arXiv},
eprint = {2312.07275},
primaryClass = {astro-ph.GA},
adsurl = {https://ui.adsabs.harvard.edu/abs/2024MNRAS.531..649G}
}

@ARTICLE{Hotan2021,
author = {{Hotan}, A.~W. and {Bunton}, J.~D. and {Chippendale}, A.~P. and {Whiting}, M. and {Tuthill}, J. and {Moss}, V.~A. and {McConnell}, D. and {Amy}, S.~W. and {Huynh}, M.~T. and {Allison}, J.~R. and {Anderson}, C.~S. and {Bannister}, K.~W. and {Bastholm}, E. and {Beresford}, R. and {Bock}, D.~C. -J. and {Bolton}, R. and {Chapman}, J.~M. and {Chow}, K. and {Collier}, J.~D. and {Cooray}, F.~R. and {Cornwell}, T.~J. and {Diamond}, P.~J. and {Edwards}, P.~G. and {Feain}, I.~J. and {Franzen}, T.~M.~O. and {George}, D. and {Gupta}, N. and {Hampson}, G.~A. and {Harvey-Smith}, L. and {Hayman}, D.~B. and {Heywood}, I. and {Jacka}, C. and {Jackson}, C.~A. and {Jackson}, S. and {Jeganathan}, K. and {Johnston}, S. and {Kesteven}, M. and {Kleiner}, D. and {Koribalski}, B.~S. and {Lee-Waddell}, K. and {Lenc}, E. and {Lensson}, E.~S. and {Mackay}, S. and {Mahony}, E.~K. and {McClure-Griffiths}, N.~M. and {McConigley}, R. and {Mirtschin}, P. and {Ng}, A.~K. and {Norris}, R.~P. and {Pearce}, S.~E. and {Phillips}, C. and {Pilawa}, M.~A. and {Raja}, W. and {Reynolds}, J.~E. and {Roberts}, P. and {Roxby}, D.~N. and {Sadler}, E.~M. and {Shields}, M. and {Schinckel}, A.~E.~T. and {Serra}, P. and {Shaw}, R.~D. and {Sweetnam}, T. and {Troup}, E.~R. and {Tzioumis}, A. and {Voronkov}, M.~A. and {Westmeier}, T.},
title = "{Australian square kilometre array pathfinder: I. system description}",
journal = {\pasa},
year = 2021,
month = mar,
volume = {38},
eid = {e009},
pages = {e009},
doi = {10.1017/pasa.2021.1},
archivePrefix = {arXiv},
eprint = {2102.01870},
primaryClass = {astro-ph.IM},
adsurl = {https://ui.adsabs.harvard.edu/abs/2021PASA...38....9H}
}

@ARTICLE{Hurley2017,
author = {{Hurley-Walker}, N. and {Callingham}, J.~R. and {Hancock}, P.~J. and {Franzen}, T.~M.~O. and {Hindson}, L. and {Kapi{\'n}ska}, A.~D. and {Morgan}, J. and {Offringa}, A.~R. and {Wayth}, R.~B. and {Wu}, C. and {Zheng}, Q. and {Murphy}, T. and {Bell}, M.~E. and {Dwarakanath}, K.~S. and {For}, B. and {Gaensler}, B.~M. and {Johnston-Hollitt}, M. and {Lenc}, E. and {Procopio}, P. and {Staveley-Smith}, L. and {Ekers}, R. and {Bowman}, J.~D. and {Briggs}, F. and {Cappallo}, R.~J. and {Deshpande}, A.~A. and {Greenhill}, L. and {Hazelton}, B.~J. and {Kaplan}, D.~L. and {Lonsdale}, C.~J. and {McWhirter}, S.~R. and {Mitchell}, D.~A. and {Morales}, M.~F. and {Morgan}, E. and {Oberoi}, D. and {Ord}, S.~M. and {Prabu}, T. and {Shankar}, N. Udaya and {Srivani}, K.~S. and {Subrahmanyan}, R. and {Tingay}, S.~J. and {Webster}, R.~L. and {Williams}, A. and {Williams}, C.~L.},
title = "{GaLactic and Extragalactic All-sky Murchison Widefield Array (GLEAM) survey - I. A low-frequency extragalactic catalogue}",
journal = {\mnras},
year = 2017,
month = jan,
volume = {464},
number = {1},
pages = {1146-1167},
doi = {10.1093/mnras/stw2337},
archivePrefix = {arXiv},
eprint = {1610.08318},
primaryClass = {astro-ph.GA},
adsurl = {https://ui.adsabs.harvard.edu/abs/2017MNRAS.464.1146H}
}

@ARTICLE{Hurley2019a,
author = {{Hurley-Walker}, N. and {Gaensler}, B.~M. and {Leahy}, D.~A. and {Filipovi{\'c}}, M.~D. and {Hancock}, P.~J. and {Franzen}, T.~M.~O. and {Offringa}, A.~R. and {Callingham}, J.~R. and {Hindson}, L. and {Wu}, C. and {Bell}, M.~E. and {For}, B. -Q. and {Johnston-Hollitt}, M. and {Kapi{\'n}ska}, A.~D. and {Morgan}, J. and {Murphy}, T. and {McKinley}, B. and {Procopio}, P. and {Staveley-Smith}, L. and {Wayth}, R.~B. and {Zheng}, Q.},
title = "{Candidate radio supernova remnants observed by the GLEAM survey over 345{\textdegree} < l < 60{\textdegree} and 180{\textdegree} < l < 240{\textdegree}}",
journal = {\pasa},
year = {2019a},
month = nov,
volume = {36},
eid = {e048},
pages = {e048},
doi = {10.1017/pasa.2019.33},
archivePrefix = {arXiv},
eprint = {1911.08124},
primaryClass = {astro-ph.HE},
adsurl = {https://ui.adsabs.harvard.edu/abs/2019PASA...36...48H}
}

@ARTICLE{Hurley2019b,
author = {{Hurley-Walker}, N. and {Filipovi{\'c}}, M.~D. and {Gaensler}, B.~M. and {Leahy}, D.~A. and {Hancock}, P.~J. and {Franzen}, T.~M.~O. and {Offringa}, A.~R. and {Callingham}, J.~R. and {Hindson}, L. and {Wu}, C. and {Bell}, M.~E. and {For}, B. -Q. and {Johnston-Hollitt}, M. and {Kapi{\'n}ska}, A.~D. and {Morgan}, J. and {Murphy}, T. and {McKinley}, B. and {Procopio}, P. and {Staveley-Smith}, L. and {Wayth}, R.~B. and {Zheng}, Q.},
title = "{New candidate radio supernova remnants detected in the GLEAM survey over 345{\textdegree} < l < 60{\textdegree}, 180{\textdegree} < l < 240{\textdegree}}",
journal = {\pasa},
year = {2019b},
month = nov,
volume = {36},
eid = {e045},
pages = {e045},
doi = {10.1017/pasa.2019.34},
archivePrefix = {arXiv},
eprint = {1911.08126},
primaryClass = {astro-ph.HE},
adsurl = {https://ui.adsabs.harvard.edu/abs/2019PASA...36...45H}
}

@ARTICLE{Hurley2019c,
author = {{Hurley-Walker}, N. and {Hancock}, P.~J. and {Franzen}, T.~M.~O. and {Callingham}, J.~R. and {Offringa}, A.~R. and {Hindson}, L. and {Wu}, C. and {Bell}, M.~E. and {For}, B. -Q. and {Gaensler}, B.~M. and {Johnston-Hollitt}, M. and {Kapi{\'n}ska}, A.~D. and {Morgan}, J. and {Murphy}, T. and {McKinley}, B. and {Procopio}, P. and {Staveley-Smith}, L. and {Wayth}, R.~B. and {Zheng}, Q.},
title = "{GaLactic and Extragalactic All-sky Murchison Widefield Array (GLEAM) survey II: Galactic plane 345{\textdegree} < l < 67{\textdegree}, 180{\textdegree} < l < 240{\textdegree}}",
journal = {\pasa},
year = {2019},
month = nov,
volume = {36},
eid = {e047},
pages = {e047},
doi = {10.1017/pasa.2019.37},
archivePrefix = {arXiv},
eprint = {1911.08127},
primaryClass = {astro-ph.GA},
adsurl = {https://ui.adsabs.harvard.edu/abs/2019PASA...36...47H}
}

@ARTICLE{Hurley2022,
author = {{Hurley-Walker}, N. and {Galvin}, T.~J. and {Duchesne}, S.~W. and {Zhang}, X. and {Morgan}, J. and {Hancock}, P.~J. and {An}, T. and {Franzen}, T.~M.~O. and {Heald}, G. and {Ross}, K. and {Vernstrom}, T. and {Anderson}, G.~E. and {Gaensler}, B.~M. and {Johnston-Hollitt}, M. and {Kaplan}, D.~L. and {Riseley}, C.~J. and {Tingay}, S.~J. and {Walker}, M.},
title = "{GaLactic and Extragalactic All-sky Murchison Widefield Array survey eXtended (GLEAM-X) I: Survey description and initial data release}",
journal = {\pasa},
year = 2022,
month = aug,
volume = {39},
eid = {e035},
pages = {e035},
doi = {10.1017/pasa.2022.17},
archivePrefix = {arXiv},
eprint = {2204.12762},
primaryClass = {astro-ph.GA},
adsurl = {https://ui.adsabs.harvard.edu/abs/2022PASA...39...35H}
}

@ARTICLE{Hurley2023,
author = {{Hurley-Walker}, N. and {Rea}, N. and {McSweeney}, S.~J. and {Meyers}, B.~W. and {Lenc}, E. and {Heywood}, I. and {Hyman}, S.~D. and {Men}, Y.~P. and {Clarke}, T.~E. and {Coti Zelati}, F. and {Price}, D.~C. and {Horv{\'a}th}, C. and {Galvin}, T.~J. and {Anderson}, G.~E. and {Bahramian}, A. and {Barr}, E.~D. and {Bhat}, N.~D.~R. and {Caleb}, M. and {Dall'Ora}, M. and {de Martino}, D. and {Giacintucci}, S. and {Morgan}, J.~S. and {Rajwade}, K.~M. and {Stappers}, B. and {Williams}, A.},
title = "{A long-period radio transient active for three decades}",
journal = {\nat},
year = 2023,
month = jul,
volume = {619},
number = {7970},
pages = {487-490},
doi = {10.1038/s41586-023-06202-5},
adsurl = {https://ui.adsabs.harvard.edu/abs/2023Natur.619..487H}
}

@INPROCEEDINGS{Kaspi1996,
author = {{Kaspi}, V.~M.},
title = "{Pulsar/Supernova Remnant Associations}",
booktitle = {IAU Colloq. 160: Pulsars: Problems and Progress},
year = 1996,
editor = {{Johnston}, S. and {Walker}, M.~A. and {Bailes}, M.},
series = {Astronomical Society of the Pacific Conference Series},
volume = {105},
month = jan,
pages = {375},
adsurl = {https://ui.adsabs.harvard.edu/abs/1996ASPC..105..375K}
}

@ARTICLE{Mainzer2011,
author = {{Mainzer}, A. and {Bauer}, J. and {Grav}, T. and {Masiero}, J. and {Cutri}, R.~M. and {Dailey}, J. and {Eisenhardt}, P. and {McMillan}, R.~S. and {Wright}, E. and {Walker}, R. and {Jedicke}, R. and {Spahr}, T. and {Tholen}, D. and {Alles}, R. and {Beck}, R. and {Brandenburg}, H. and {Conrow}, T. and {Evans}, T. and {Fowler}, J. and {Jarrett}, T. and {Marsh}, K. and {Masci}, F. and {McCallon}, H. and {Wheelock}, S. and {Wittman}, M. and {Wyatt}, P. and {DeBaun}, E. and {Elliott}, G. and {Elsbury}, D. and {Gautier}, T., IV and {Gomillion}, S. and {Leisawitz}, D. and {Maleszewski}, C. and {Micheli}, M. and {Wilkins}, A.},
title = "{Preliminary Results from NEOWISE: An Enhancement to the Wide-field Infrared Survey Explorer for Solar System Science}",
journal = {\apj},
year = 2011,
month = apr,
volume = {731},
number = {1},
eid = {53},
pages = {53},
doi = {10.1088/0004-637X/731/1/53},
archivePrefix = {arXiv},
eprint = {1102.1996},
primaryClass = {astro-ph.EP},
adsurl = {https://ui.adsabs.harvard.edu/abs/2011ApJ...731...53M}
}

@article{Manchester2005,
doi = {10.1086/428488},
url = {https://dx.doi.org/10.1086/428488},
year = {2005},
month = {04},
volume = {129},
number = {4},
pages = {1993},
author = {R. N. Manchester and G. B. Hobbs and A. Teoh and M. Hobbs},
title = {The Australia Telescope National Facility Pulsar Catalogue},
journal = {The Astronomical Journal}
}

@ARTICLE{Mantovanini2024,
author = {{Mantovanini}, S. and {Becker}, W. and {Khokhriakova}, A. and {Hurley-Walker}, N. and {Anderson}, G.~E. and {Nicastro}, L.},
title = "{G321.3-3.9: a new supernova remnant observed with multi-band radio data and in the SRG/eROSITA All-Sky Surveys}",
journal = {arXiv e-prints},
year = 2024,
month = jan,
eid = {arXiv:2401.17294},
pages = {arXiv:2401.17294},
doi = {10.48550/arXiv.2401.17294},
archivePrefix = {arXiv},
eprint = {2401.17294},
primaryClass = {astro-ph.HE},
adsurl = {https://ui.adsabs.harvard.edu/abs/2024arXiv240117294M}
}

@ARTICLE{McClureGriffiths2001,
author = {{McClure-Griffiths}, N.~M. and {Green}, A.~J. and {Dickey}, John M. and {Gaensler}, B.~M. and {Haynes}, R.~F. and {Wieringa}, M.~H.},
title = "{The Southern Galactic Plane Survey: The Test Region}",
journal = {\apj},
year = 2001,
month = apr,
volume = {551},
number = {1},
pages = {394-412},
doi = {10.1086/320095},
archivePrefix = {arXiv},
eprint = {astro-ph/0012302},
primaryClass = {astro-ph},
adsurl = {https://ui.adsabs.harvard.edu/abs/2001ApJ...551..394M}
}

@ARTICLE{Mcconnell2020,
author = {{McConnell}, D. and {Hale}, C.~L. and {Lenc}, E. and {Banfield}, J.~K. and {Heald}, George and {Hotan}, A.~W. and {Leung}, James K. and {Moss}, Vanessa A. and {Murphy}, Tara and {O'Brien}, Andrew and {Pritchard}, Joshua and {Raja}, Wasim and {Sadler}, Elaine M. and {Stewart}, Adam and {Thomson}, Alec J.~M. and {Whiting}, M. and {Allison}, James R. and {Amy}, S.~W. and {Anderson}, C. and {Ball}, Lewis and {Bannister}, Keith W. and {Bell}, Martin and {Bock}, Douglas C. -J. and {Bolton}, Russ and {Bunton}, J.~D. and {Chippendale}, A.~P. and {Collier}, J.~D. and {Cooray}, F.~R. and {Cornwell}, T.~J. and {Diamond}, P.~J. and {Edwards}, P.~G. and {Gupta}, N. and {Hayman}, Douglas B. and {Heywood}, Ian and {Jackson}, C.~A. and {Koribalski}, B{\"a}rbel S. and {Lee-Waddell}, Karen and {McClure-Griffiths}, N.~M. and {Ng}, Alan and {Norris}, Ray P. and {Phillips}, Chris and {Reynolds}, John E. and {Roxby}, Daniel N. and {Schinckel}, Antony E.~T. and {Shields}, Matt and {Tremblay}, Chenoa and {Tzioumis}, A. and {Voronkov}, M.~A. and {Westmeier}, Tobias},
title = "{The Rapid ASKAP Continuum Survey I: Design and first results}",
journal = {\pasa},
year = 2020,
month = nov,
volume = {37},
eid = {e048},
pages = {e048},
doi = {10.1017/pasa.2020.41},
archivePrefix = {arXiv},
eprint = {2012.00747},
primaryClass = {astro-ph.IM},
adsurl = {https://ui.adsabs.harvard.edu/abs/2020PASA...37...48M}
}

@ARTICLE{Norris2021,
author = {{Norris}, Ray P. and {Marvil}, Joshua and {Collier}, J.~D. and {Kapi{\'n}ska}, Anna D. and {O'Brien}, Andrew N. and {Rudnick}, L. and {Andernach}, Heinz and {Asorey}, Jacobo and {Brown}, Michael J.~I. and {Br{\"u}ggen}, Marcus and {Crawford}, Evan and {English}, Jayanne and {Rahman}, Syed Faisal ur and {Filipovi{\'c}}, Miroslav D. and {Gordon}, Yjan and {G{\"u}rkan}, G{\"u}lay and {Hale}, Catherine and {Hopkins}, Andrew M. and {Huynh}, Minh T. and {HyeongHan}, Kim and {James Jee}, M. and {Koribalski}, B{\"a}rbel S. and {Lenc}, Emil and {Luken}, Kieran and {Parkinson}, David and {Prandoni}, Isabella and {Raja}, Wasim and {Reiprich}, Thomas H. and {Riseley}, Christopher J. and {Shabala}, Stanislav S. and {Sheil}, Jaimie R. and {Vernstrom}, Tessa and {Whiting}, Matthew T. and {Allison}, James R. and {Anderson}, C.~S. and {Ball}, Lewis and {Bell}, Martin and {Bunton}, John and {Galvin}, T.~J. and {Gupta}, Neeraj and {Hotan}, Aidan and {Jacka}, Colin and {Macgregor}, Peter J. and {Mahony}, Elizabeth K. and {Maio}, Umberto and {Moss}, Vanessa and {Pandey-Pommier}, M. and {Voronkov}, Maxim A.},
title = "{The Evolutionary Map of the Universe pilot survey}",
journal = {\pasa},
year = 2021,
month = sep,
volume = {38},
eid = {e046},
pages = {e046},
doi = {10.1017/pasa.2021.42},
archivePrefix = {arXiv},
eprint = {2108.00569},
primaryClass = {astro-ph.CO},
adsurl = {https://ui.adsabs.harvard.edu/abs/2021PASA...38...46N}
}

@ARTICLE{Offringa2012,
author = {{Offringa}, A.~R. and {van de Gronde}, J.~J. and {Roerdink}, J.~B.~T.~M.},
title = "{A morphological algorithm for improving radio-frequency interference detection}",
journal = {\aap},
year = 2012,
month = mar,
volume = {539},
eid = {A95},
pages = {A95},
doi = {10.1051/0004-6361/201118497},
archivePrefix = {arXiv},
eprint = {1201.3364},
primaryClass = {astro-ph.IM},
adsurl = {https://ui.adsabs.harvard.edu/abs/2012A&A...539A..95O}
}

@ARTICLE{Offringa2014,
author = {{Offringa}, A.~R. and {McKinley}, B. and {Hurley-Walker}, N. and {Briggs}, F.~H. and {Wayth}, R.~B. and {Kaplan}, D.~L. and {Bell}, M.~E. and {Feng}, L. and {Neben}, A.~R. and {Hughes}, J.~D. and {Rhee}, J. and {Murphy}, T. and {Bhat}, N.~D.~R. and {Bernardi}, G. and {Bowman}, J.~D. and {Cappallo}, R.~J. and {Corey}, B.~E. and {Deshpande}, A.~A. and {Emrich}, D. and {Ewall-Wice}, A. and {Gaensler}, B.~M. and {Goeke}, R. and {Greenhill}, L.~J. and {Hazelton}, B.~J. and {Hindson}, L. and {Johnston-Hollitt}, M. and {Jacobs}, D.~C. and {Kasper}, J.~C. and {Kratzenberg}, E. and {Lenc}, E. and {Lonsdale}, C.~J. and {Lynch}, M.~J. and {McWhirter}, S.~R. and {Mitchell}, D.~A. and {Morales}, M.~F. and {Morgan}, E. and {Kudryavtseva}, N. and {Oberoi}, D. and {Ord}, S.~M. and {Pindor}, B. and {Procopio}, P. and {Prabu}, T. and {Riding}, J. and {Roshi}, D.~A. and {Shankar}, N. Udaya and {Srivani}, K.~S. and {Subrahmanyan}, R. and {Tingay}, S.~J. and {Waterson}, M. and {Webster}, R.~L. and {Whitney}, A.~R. and {Williams}, A. and {Williams}, C.~L.},
title = "{WSCLEAN: an implementation of a fast, generic wide-field imager for radio astronomy}",
journal = {\mnras},
year = 2014,
month = oct,
volume = {444},
number = {1},
pages = {606-619},
doi = {10.1093/mnras/stu1368},
archivePrefix = {arXiv},
eprint = {1407.1943},
primaryClass = {astro-ph.IM},
adsurl = {https://ui.adsabs.harvard.edu/abs/2014MNRAS.444..606O}
}

@ARTICLE{Olausen2014,
author = {{Olausen}, S.~A. and {Kaspi}, V.~M.},
title = "{The McGill Magnetar Catalog}",
journal = {\apjs},
year = 2014,
month = may,
volume = {212},
number = {1},
eid = {6},
pages = {6},
doi = {10.1088/0067-0049/212/1/6},
archivePrefix = {arXiv},
eprint = {1309.4167},
primaryClass = {astro-ph.HE},
adsurl = {https://ui.adsabs.harvard.edu/abs/2014ApJS..212....6O}
}

@ARTICLE{Predehl2021,
author = {{Predehl}, P. and {Andritschke}, R. and {Arefiev}, V. and {Babyshkin}, V. and {Batanov}, O. and {Becker}, W. and {B{\"o}hringer}, H. and {Bogomolov}, A. and {Boller}, T. and {Borm}, K. and {Bornemann}, W. and {Br{\"a}uninger}, H. and {Br{\"u}ggen}, M. and {Brunner}, H. and {Brusa}, M. and {Bulbul}, E. and {Buntov}, M. and {Burwitz}, V. and {Burkert}, W. and {Clerc}, N. and {Churazov}, E. and {Coutinho}, D. and {Dauser}, T. and {Dennerl}, K. and {Doroshenko}, V. and {Eder}, J. and {Emberger}, V. and {Eraerds}, T. and {Finoguenov}, A. and {Freyberg}, M. and {Friedrich}, P. and {Friedrich}, S. and {F{\"u}rmetz}, M. and {Georgakakis}, A. and {Gilfanov}, M. and {Granato}, S. and {Grossberger}, C. and {Gueguen}, A. and {Gureev}, P. and {Haberl}, F. and {H{\"a}lker}, O. and {Hartner}, G. and {Hasinger}, G. and {Huber}, H. and {Ji}, L. and {Kienlin}, A. v. and {Kink}, W. and {Korotkov}, F. and {Kreykenbohm}, I. and {Lamer}, G. and {Lomakin}, I. and {Lapshov}, I. and {Liu}, T. and {Maitra}, C. and {Meidinger}, N. and {Menz}, B. and {Merloni}, A. and {Mernik}, T. and {Mican}, B. and {Mohr}, J. and {M{\"u}ller}, S. and {Nandra}, K. and {Nazarov}, V. and {Pacaud}, F. and {Pavlinsky}, M. and {Perinati}, E. and {Pfeffermann}, E. and {Pietschner}, D. and {Ramos-Ceja}, M.~E. and {Rau}, A. and {Reiffers}, J. and {Reiprich}, T.~H. and {Robrade}, J. and {Salvato}, M. and {Sanders}, J. and {Santangelo}, A. and {Sasaki}, M. and {Scheuerle}, H. and {Schmid}, C. and {Schmitt}, J. and {Schwope}, A. and {Shirshakov}, A. and {Steinmetz}, M. and {Stewart}, I. and {Str{\"u}der}, L. and {Sunyaev}, R. and {Tenzer}, C. and {Tiedemann}, L. and {Tr{\"u}mper}, J. and {Voron}, V. and {Weber}, P. and {Wilms}, J. and {Yaroshenko}, V.},
title = "{The eROSITA X-ray telescope on SRG}",
journal = {\aap},
year = 2021,
month = mar,
volume = {647},
eid = {A1},
pages = {A1},
doi = {10.1051/0004-6361/202039313},
archivePrefix = {arXiv},
eprint = {2010.03477},
primaryClass = {astro-ph.HE},
adsurl = {https://ui.adsabs.harvard.edu/abs/2021A&A...647A...1P}
}

@ARTICLE{Ranasinghe2023,
author = {{Ranasinghe}, S. and {Leahy}, D.},
title = "{A Statistical Analysis of Galactic Radio Supernova Remnants}",
journal = {\apjs},
year = 2023,
month = apr,
volume = {265},
number = {2},
eid = {53},
pages = {53},
doi = {10.3847/1538-4365/acc1de},
archivePrefix = {arXiv},
eprint = {2302.06593},
primaryClass = {astro-ph.GA},
adsurl = {https://ui.adsabs.harvard.edu/abs/2023ApJS..265...53R}
}

@ARTICLE{Reynolds1994,
author = {{Reynolds}, S.~P. and {Lyutikov}, M. and {Blandford}, R.~D. and {Seward}, F.~D.},
title = "{X-ray evidence for the association of G11.2-0.3 with the supernova of 386 AD.}",
journal = {\mnras},
year = 1994,
month = nov,
volume = {271},
pages = {L1-L4},
doi = {10.1093/mnras/271.1.L1},
adsurl = {https://ui.adsabs.harvard.edu/abs/1994MNRAS.271L...1R}
}

@ARTICLE{Ross2024,
author = {{Ross}, K. and {Hurley-Walker}, N. and {Galvin}, T.~J. and {Venville}, B. and {Duchesne}, S.~W. and {Morgan}, J. and {An}, T. and {Gurkan}, G. and {Hancock}, P.~J. and {Heald}, G. and {Johnston-Hollitt}, M. and {White}, S.~V.},
title = "{GaLactic and Extragalactic All-sky Murchison Widefield Array eXtended (GLEAM-X) survey II: Second Data Release}",
journal = {arXiv e-prints},
year = 2024,
month = jun,
eid = {arXiv:2406.06921},
pages = {arXiv:2406.06921},
doi = {10.48550/arXiv.2406.06921},
archivePrefix = {arXiv},
eprint = {2406.06921},
primaryClass = {astro-ph.GA},
adsurl = {https://ui.adsabs.harvard.edu/abs/2024arXiv240606921R}
}

@ARTICLE{Stephenson1971,
author = {{Stephenson}, F.~R.},
title = "{Suspected Supernova in A.D. 1181}",
journal = {\qjras},
year = 1971,
month = mar,
volume = {12},
pages = {10},
adsurl = {https://ui.adsabs.harvard.edu/abs/1971QJRAS..12...10S}
}

@ARTICLE{Stephenson1977,
author = {{Stephenson}, F. Richard and {Clark}, David H. and {Crawford}, David F.},
title = "{The supernova of AD 1006}",
journal = {\mnras},
year = 1977,
month = sep,
volume = {180},
pages = {567-584},
doi = {10.1093/mnras/180.4.567},
adsurl = {https://ui.adsabs.harvard.edu/abs/1977MNRAS.180..567S}
}

@ARTICLE{Tammann1994,
author = {{Tammann}, G.~A. and {Loeffler}, W. and {Schroeder}, A.},
title = "{The Galactic Supernova Rate}",
journal = {\apjs},
year = 1994,
month = jun,
volume = {92},
pages = {487},
doi = {10.1086/192002},
adsurl = {https://ui.adsabs.harvard.edu/abs/1994ApJS...92..487T}
}

@ARTICLE{Tingay2013,
author = {{Tingay}, S.~J. and {Goeke}, R. and {Bowman}, J.~D. and {Emrich}, D. and {Ord}, S.~M. and {Mitchell}, D.~A. and {Morales}, M.~F. and {Booler}, T. and {Crosse}, B. and {Wayth}, R.~B. and {Lonsdale}, C.~J. and {Tremblay}, S. and {Pallot}, D. and {Colegate}, T. and {Wicenec}, A. and {Kudryavtseva}, N. and {Arcus}, W. and {Barnes}, D. and {Bernardi}, G. and {Briggs}, F. and {Burns}, S. and {Bunton}, J.~D. and {Cappallo}, R.~J. and {Corey}, B.~E. and {Deshpande}, A. and {Desouza}, L. and {Gaensler}, B.~M. and {Greenhill}, L.~J. and {Hall}, P.~J. and {Hazelton}, B.~J. and {Herne}, D. and {Hewitt}, J.~N. and {Johnston-Hollitt}, M. and {Kaplan}, D.~L. and {Kasper}, J.~C. and {Kincaid}, B.~B. and {Koenig}, R. and {Kratzenberg}, E. and {Lynch}, M.~J. and {Mckinley}, B. and {Mcwhirter}, S.~R. and {Morgan}, E. and {Oberoi}, D. and {Pathikulangara}, J. and {Prabu}, T. and {Remillard}, R.~A. and {Rogers}, A.~E.~E. and {Roshi}, A. and {Salah}, J.~E. and {Sault}, R.~J. and {Udaya-Shankar}, N. and {Schlagenhaufer}, F. and {Srivani}, K.~S. and {Stevens}, J. and {Subrahmanyan}, R. and {Waterson}, M. and {Webster}, R.~L. and {Whitney}, A.~R. and {Williams}, A. and {Williams}, C.~L. and {Wyithe}, J.~S.~B.},
title = "{The Murchison Widefield Array: The Square Kilometre Array Precursor at Low Radio Frequencies}",
journal = {\pasa},
year = 2013,
month = jan,
volume = {30},
eid = {e007},
pages = {e007},
doi = {10.1017/pasa.2012.007},
archivePrefix = {arXiv},
eprint = {1206.6945},
primaryClass = {astro-ph.IM},
adsurl = {https://ui.adsabs.harvard.edu/abs/2013PASA...30....7T}
}

@ARTICLE{VandenBergh1977,
author = {{van den Bergh}, S. and {Kamper}, K.~W.},
title = "{The remnant of Kepler's supernova.}",
journal = {\apj},
year = 1977,
month = dec,
volume = {218},
pages = {617},
doi = {10.1086/155719},
adsurl = {https://ui.adsabs.harvard.edu/abs/1977ApJ...218..617V}
}

@ARTICLE{Watson2008,
author = {{Watson}, C. and {Povich}, M.~S. and {Churchwell}, E.~B. and {Babler}, B.~L. and {Chunev}, G. and {Hoare}, M. and {Indebetouw}, R. and {Meade}, M.~R. and {Robitaille}, T.~P. and {Whitney}, B.~A.},
title = "{Infrared Dust Bubbles: Probing the Detailed Structure and Young Massive Stellar Populations of Galactic H II Regions}",
journal = {\apj},
year = 2008,
month = jul,
volume = {681},
number = {2},
pages = {1341-1355},
doi = {10.1086/588005},
archivePrefix = {arXiv},
eprint = {0806.0609},
primaryClass = {astro-ph},
adsurl = {https://ui.adsabs.harvard.edu/abs/2008ApJ...681.1341W}
}

@ARTICLE{Wayth2015,
author = {{Wayth}, R.~B. and {Lenc}, E. and {Bell}, M.~E. and {Callingham}, J.~R. and {Dwarakanath}, K.~S. and {Franzen}, T.~M.~O. and {For}, B. -Q. and {Gaensler}, B. and {Hancock}, P. and {Hindson}, L. and {Hurley-Walker}, N. and {Jackson}, C.~A. and {Johnston-Hollitt}, M. and {Kapi{\'n}ska}, A.~D. and {McKinley}, B. and {Morgan}, J. and {Offringa}, A.~R. and {Procopio}, P. and {Staveley-Smith}, L. and {Wu}, C. and {Zheng}, Q. and {Trott}, C.~M. and {Bernardi}, G. and {Bowman}, J.~D. and {Briggs}, F. and {Cappallo}, R.~J. and {Corey}, B.~E. and {Deshpande}, A.~A. and {Emrich}, D. and {Goeke}, R. and {Greenhill}, L.~J. and {Hazelton}, B.~J. and {Kaplan}, D.~L. and {Kasper}, J.~C. and {Kratzenberg}, E. and {Lonsdale}, C.~J. and {Lynch}, M.~J. and {McWhirter}, S.~R. and {Mitchell}, D.~A. and {Morales}, M.~F. and {Morgan}, E. and {Oberoi}, D. and {Ord}, S.~M. and {Prabu}, T. and {Rogers}, A.~E.~E. and {Roshi}, A. and {Shankar}, N. Udaya and {Srivani}, K.~S. and {Subrahmanyan}, R. and {Tingay}, S.~J. and {Waterson}, M. and {Webster}, R.~L. and {Whitney}, A.~R. and {Williams}, A. and {Williams}, C.~L.},
title = "{GLEAM: The GaLactic and Extragalactic All-Sky MWA Survey}",
journal = {\pasa},
year = 2015,
month = jun,
volume = {32},
eid = {e025},
pages = {e025},
doi = {10.1017/pasa.2015.26},
archivePrefix = {arXiv},
eprint = {1505.06041},
primaryClass = {astro-ph.IM},
adsurl = {https://ui.adsabs.harvard.edu/abs/2015PASA...32...25W}
}

@ARTICLE{Wayth2018,
author = {{Wayth}, Randall B. and {Tingay}, Steven J. and {Trott}, Cathryn M. and {Emrich}, David and {Johnston-Hollitt}, Melanie and {McKinley}, Ben and {Gaensler}, B.~M. and {Beardsley}, A.~P. and {Booler}, T. and {Crosse}, B. and {Franzen}, T.~M.~O. and {Horsley}, L. and {Kaplan}, D.~L. and {Kenney}, D. and {Morales}, M.~F. and {Pallot}, D. and {Sleap}, G. and {Steele}, K. and {Walker}, M. and {Williams}, A. and {Wu}, C. and {Cairns}, Iver. H. and {Filipovic}, M.~D. and {Johnston}, S. and {Murphy}, T. and {Quinn}, P. and {Staveley-Smith}, L. and {Webster}, R. and {Wyithe}, J.~S.~B.},
title = "{The Phase II Murchison Widefield Array: Design overview}",
journal = {\pasa},
year = 2018,
month = nov,
volume = {35},
eid = {e033},
pages = {e033},
doi = {10.1017/pasa.2018.37},
archivePrefix = {arXiv},
eprint = {1809.06466},
primaryClass = {astro-ph.IM},
adsurl = {https://ui.adsabs.harvard.edu/abs/2018PASA...35...33W}
}

@ARTICLE{Whiteoak1996,
author = {{Whiteoak}, J.~B.~Z. and {Green}, A.~J.},
title = "{The MOST supernova remnant catalogue (MSC).}",
journal = {\aaps},
year = 1996,
month = aug,
volume = {118},
pages = {329-380},
adsurl = {https://ui.adsabs.harvard.edu/abs/1996A&AS..118..329W}
}

@ARTICLE{Wright2010,
author = {{Wright}, Edward L. and {Eisenhardt}, Peter R.~M. and {Mainzer}, Amy K. and {Ressler}, Michael E. and {Cutri}, Roc M. and {Jarrett}, Thomas and {Kirkpatrick}, J. Davy and {Padgett}, Deborah and {McMillan}, Robert S. and {Skrutskie}, Michael and {Stanford}, S.~A. and {Cohen}, Martin and {Walker}, Russell G. and {Mather}, John C. and {Leisawitz}, David and {Gautier}, Thomas N., III and {McLean}, Ian and {Benford}, Dominic and {Lonsdale}, Carol J. and {Blain}, Andrew and {Mendez}, Bryan and {Irace}, William R. and {Duval}, Valerie and {Liu}, Fengchuan and {Royer}, Don and {Heinrichsen}, Ingolf and {Howard}, Joan and {Shannon}, Mark and {Kendall}, Martha and {Walsh}, Amy L. and {Larsen}, Mark and {Cardon}, Joel G. and {Schick}, Scott and {Schwalm}, Mark and {Abid}, Mohamed and {Fabinsky}, Beth and {Naes}, Larry and {Tsai}, Chao-Wei},
title = "{The Wide-field Infrared Survey Explorer (WISE): Mission Description and Initial On-orbit Performance}",
journal = {\aj},
year = 2010,
month = dec,
volume = {140},
number = {6},
pages = {1868-1881},
doi = {10.1088/0004-6256/140/6/1868},
archivePrefix = {arXiv},
eprint = {1008.0031},
primaryClass = {astro-ph.IM},
adsurl = {https://ui.adsabs.harvard.edu/abs/2010AJ....140.1868W}
}

\end{document}